\documentclass[aps,prx,reprint,amsfonts,amsmath,amssymb,superscriptaddress,longbibliography,floatfix]{revtex4-2}
\usepackage{comment} 
\usepackage[T1]{fontenc}
\usepackage{graphicx}
\usepackage[dvipsnames]{xcolor}
\usepackage{braket}
\usepackage{enumitem}
\usepackage{pifont}
\usepackage[english]{babel}
\usepackage{titlesec} 
\usepackage[colorlinks,linkcolor=Blue,citecolor=Blue]{hyperref}
\usepackage{tikz}
\usepackage{circuitikz}
\usetikzlibrary{positioning}
\usepackage{multirow}
\usepackage{soul}
\usepackage{cancel}

\begin{document}
\title{Berry curvature dipole circuit quantum electrodynamics}
\author{Oles Matsyshyn}
\affiliation{Division of Physics and Applied Physics, School of Physical and Mathematical Sciences, Nanyang Technological University, Singapore 637371}
\author{Valla Fatemi}
\email{vf82@cornell.edu}
\affiliation{School of Applied and Engineering Physics, Cornell University, Ithaca, NY 14853, USA}
\author{Justin C.W. Song}
\email{justinsong@ntu.edu.sg}
\affiliation{Division of Physics and Applied Physics, School of Physical and Mathematical Sciences, Nanyang Technological University, Singapore 637371}
\date{\today}
 
\begin{abstract}
Quantum three-wave mixing is an important resource in superconducting quantum circuits. The traditional implementation uses Josephson junction arrays and is activated by magnetic flux.  Here we introduce the Berry curvature dipole (BCD) quantum circuit element, which achieves quantum three-wave mixing without breaking time-reversal symmetry. This quantum three-wave mixing results from the BCD nonlinear Hall effect in noncentrosymmetric quantum materials taken to a dissipationless, quantum coherent few-photon regime. We show how nonlinear Hall materials can be integrated to superconducting microwave circuits to accomplish multiphoton vacuum Rabi splittings and dynamical stabilization of bosonic Schr\"odinger cat states. Such BCD circuit quantum electrodynamics are a concrete exemplar of a physical resource for quantum information processing enabled by nonlinear response in quantum materials.
\end{abstract}

\maketitle

Dissipationless nonlinearities play an essential role in circuit quantum electrodynamics~\cite{Rev}. Such quantum nonlinearities enable anharmonic energy levels critical for readily addressable superconducting qubit states~\cite{Clarke,Koch-Transmon,Rev}, control parametric processes useful in entanglement generation~\cite{storz_loophole-free_2023}, and underlie quantum sensing strategies~\cite{Fong}. At its core are Josephson junctions~\cite{Josephson} --- the key workhorse circuit element for implementing quantum coherent nonlinearities in superconducting quantum circuits. Formed from weak links between superconducting electrodes, Josephson junction nonlinearities originate from Cooper pair tunneling across the junction~\cite{Tinkham}. As such, Josephson junction circuit engineering has dominated architectures for quantum information processing and devices~\cite{Rev}.

\begin{figure}[t]
    \centering
    \includegraphics[width=0.5\textwidth]{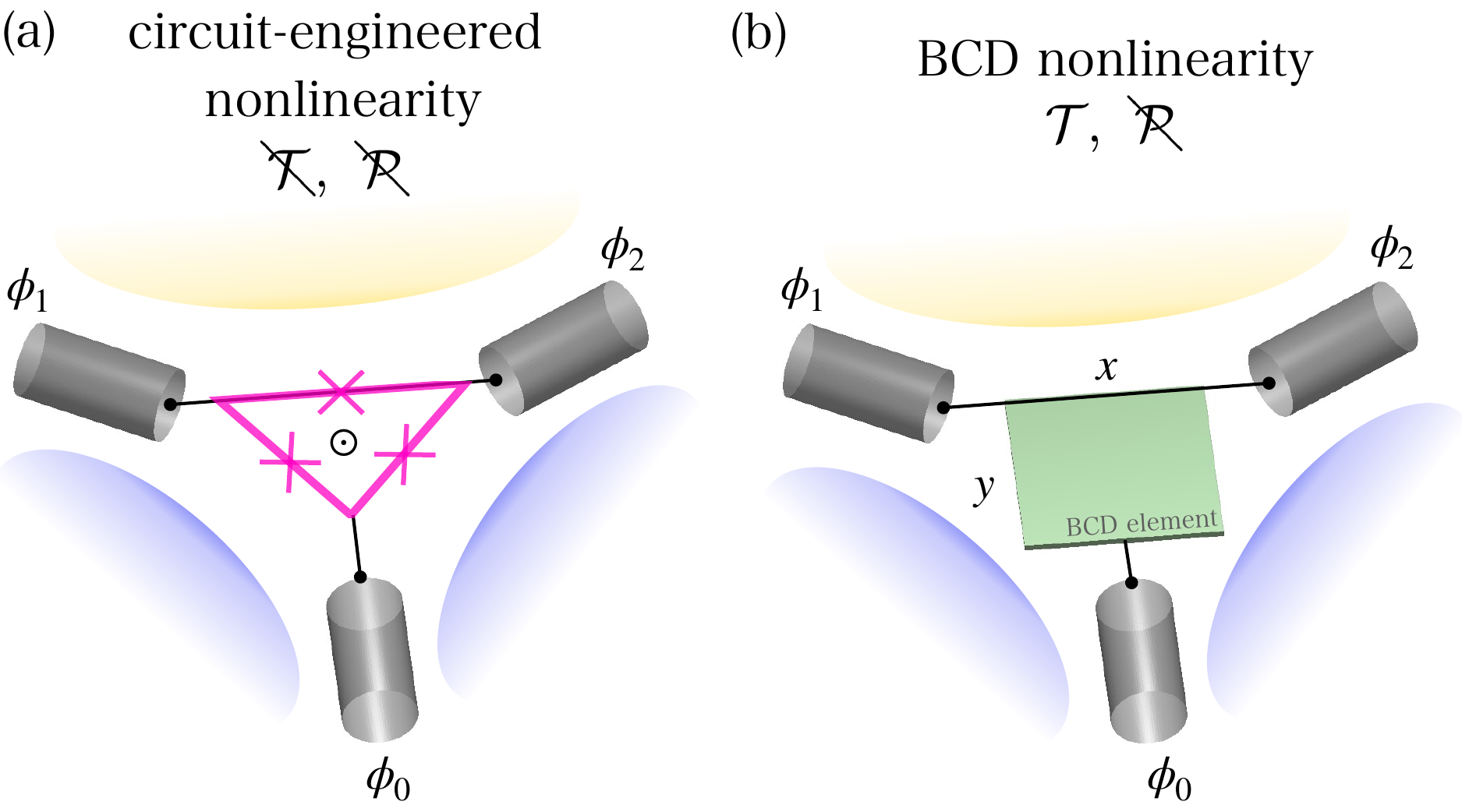}
    \caption{\textbf{Circuit-engineered versus BCD quantum three-wave mixing nonlinearity.} (a) Quantum three-wave mixing can be engineered in a Josephson junction based circuit that requires both broken $\mathcal{T}$ and $\mathcal{P}$. In these, $\mathcal{T}$ breaking is controlled by applied magnetic flux $\phi_{\rm ext}$. This panel shows a three-terminal Josephson junction based device. In contrast, noncentrosymmetric superconductors can realize quantum three-wave mixing intrinsically without $\mathcal{T}$ breaking via a (b) BCD nonlinear quantum circuit element. Here green represents a non-centrosymmetric superconductor that mediates a nonlinear Hall response. This BCD circuit element naturally couples dynamics along the $x$ and $y$ directions. In both panels, $\phi_{1,2}$ are phases and $\phi_0$ is a reference phase.}\label{sBCDQFIG1}
\end{figure}

On its own, the Josephson junction preserves both time-reversal symmetry $\mathcal{T}$ and inversion symmetry $\mathcal{P}$; these restrict its potential energy to be symmetric in the superconducting phase, constraining the nonlinear dynamics to offer only even-order wave mixing effects (e.g., four-wave mixing). To engineer responses beyond these constraints, symmetries are broken at the circuit level. For instance, to achieve quantum three-wave mixing processes, multiple Josephson junctions in a loop with applied magnetic flux are often employed to break $\mathcal{T}$ and $\mathcal{P}$ together \cite{Fratinni-Snail,autocat} (e.g., Fig.~\ref{sBCDQFIG1}(a)). Similarly, three-wave mixing can also be achieved in noncentrosymmetric superconducting magnets where $\mathcal{T}$ and $\mathcal{P}$ are broken intrinsically in the constituent materials \cite{Schrade-Fatemi}. In all these, breaking $\mathcal{T}$ has been a keystone of symmetry-broken Josephson array design and nonlinear superconducting circuit physics. 

Here we introduce a quantum materials paradigm for dissipationless nonlinearities in $\mathcal{T}$-preserving superconducting quantum circuits. Specifically, we propose a Berry curvature dipole (BCD) nonlinear quantum circuit element (Fig. \ref{sBCDQFIG1}(b)) that can be built from noncentrosymmetric superconductors. The BCD nonlinear quantum circuit element enables {\it dissipationless} three-wave mixing, $g_3$, with $\mathcal{T}$ symmetry intact. 
This BCD mixing arises from intrinsic nonlinear Hall currents in bulk noncentrosymmetric superconductors~\cite{sBCD_accepted, Yanase1, Yanase2}, which we argue can operate in a circuit at the few-photon quantum limit. This element possesses a fundamentally different circuit Lagrangian structure to that of three-wave mixing Josephson junction devices that break $\mathcal{T}$. As a result, the BCD nonlinear quantum circuit element produces a distinct BCD-induced circuit quantum electrodynamics. 

We predict BCD circuit Hamiltonian terms $g_3$ at the MHz scale for readily available noncentrosymmetric superconductor platforms. This offers both a novel route to investigating the superconducting BCD effect itself and a nonlinearity for driven-dissipative protocols~\cite{MirrahimiIOP_2014,autocat} for defining bosonic quantum information systems. Together, these render the BCD nonlinear quantum circuit and noncentrosymmetric quantum materials a novel platform and design element for realizing magnetism-free nonlinear superconducting quantum devices. 

\vspace{2mm}
\textit{\textcolor{blue}{Bulk superconducting nonlinearities.}} We begin by examining the current response in a gapped superconductor from a general symmetry perspective. For operating frequencies, $\omega$, smaller than the superconducting optical gap, $\omega < 2\Delta_{\rm sc}/\hbar$, the supercurrent is dissipationless and directly depends on condensate momentum $2\hbar \mathbf{q}$, $\mathbf{q} (t)= \mathbf{q}_{0} +(\mathbf{q}_\omega e^{-i\omega t} + \mathbf{q}_{-\omega} e^{i\omega t})$~\cite{sBCD_accepted}. Here the oscillating component can be produced by an oscillating vector potential (or oscillating electric field) found in quantum circuits. This correspondence enables to directly track the dynamical supercurrent $\mathbf{j}(t)$ in powers of $\mathbf{q} (t)$:
\begin{multline}
\label{eq:currentfull}
    j^\gamma \!=\! \frac{e}{\hbar}\rho^{\gamma \alpha} q^\alpha \!+\!  \chi^{\gamma\alpha\beta} q^\alpha q^\beta\!+\! \tilde{\chi}^{\gamma\alpha\beta} q^\alpha \dot q^\beta\!+\!  \mathcal{O}(q^3\!\!,\omega^2),
\end{multline}
where $\rho$ is the familiar superfluid stiffness tensor that describes conventional supercurrent flow linear in $\mathbf{q}$, and $t$-dependence is implicit for notational brevity. In contrast, $\chi$ and $\tilde{\chi}$ characterize a supercurrent flow that is nonlinear in $\mathbf{q}$. The second-order nonlinear nature of  $\chi$ and $\tilde{\chi}$ means that both require broken centrosymmetry. The two terms are differentiated by time-reversal: $\chi$ ($\tilde{\chi}$) is time-reversal odd (even). As a result, $\chi$ vanishes in noncentrosymmetric superconductors without magnetism, whereas $\tilde{\chi}$ persists in non-magnetic noncentrosymmetric superconductors producing a nonlinear supercurrent controlled by $q^\alpha\dot{q}^\beta$. In what follows, we will focus on the $\tilde{\chi}$ nonlinearity that manifest in noncentrosymmetric superconductors with no magnetic fields. 

Interestingly, the dissipationless nature of supercurrent imposes strong constraints on the allowed components of $\tilde{\chi}$ in a noncentrosymmetric superconductor. To see this, consider the net energy absorption per oscillation cycle $P = \langle j^\gamma E^\gamma \rangle_{\rm cyc} = (-\hbar/e)\langle j^\gamma \dot{q}^\gamma\rangle_{\rm cyc}$, where $\hbar \dot{q}^\gamma = -e E^\gamma$ for an AC electric field; here the cycle average is $\langle O(t) \rangle_{\rm cyc} = (\omega/2\pi)\int_0^{2\pi/\omega}O(t) dt$. For frequencies well below the superconducting gap and at zero temperature, $P$ vanishes due to the non-dissipative nature of supercurrent flow. This is naturally satisfied by the familiar superfluid stiffness current $\rho^{\gamma \alpha} q^\alpha$ since $2P_\rho = \langle d[\rho^{\gamma\alpha}q^\alpha q^\gamma]/dt \rangle_{\rm cyc} =0$. Moving to $\tilde{\chi}$ induced nonlinear currents, we find $P_{\tilde{\chi}} = P_{+} + P_-$ where $P_+ = \langle \tilde{\chi}_{+}^{\gamma\alpha \beta} q^\alpha \dot{q}^\beta \dot{q}^\gamma\rangle_{\rm cyc}$ and $P_- = \langle \tilde{\chi}_{-}^{\gamma\alpha \beta} q^\alpha \dot{q}^\beta \dot{q}^\gamma\rangle_{\rm cyc}$; here $\tilde{\chi}_{\pm}$ is the symmetric (anti-symmetric) part of $\tilde{\chi}$ in ($\beta,\gamma$). 

Notice that $P_+ = 2\omega^2 \tilde{\chi}_{+}^{\gamma\alpha \beta}q_0^{\alpha}{\rm Re}[q_{\omega}^\beta q_{-\omega}^{\gamma}]$. Our operating regime $\omega < 2\Delta_{\rm sc}/\hbar$ enforces $P=0$, which means that the symmetric $\tilde{\chi}_{+} =0$ must vanish. In contrast, evaluating the antisymmetric component, we find $P_- =0$ for nonzero values of the response because $\tilde{\chi}^{\gamma\alpha\beta}_-=-\tilde{\chi}^{\beta\alpha\gamma}_-$. As a result, $\tilde{\chi}^{\gamma\alpha\beta}_-$ is a finite {\it dissipationless} nonlinearity, which can intrinsically manifest in a superconductor. 

The antisymmetric nature of $\tilde{\chi}^{\gamma\alpha\beta}_-$ generates nonlinear supercurrents transverse to $\dot{\mathbf{q}}$: a superconducting analogue of a \textit{nonlinear} Hall effect~\cite{sBCD_accepted,SodemannFu,Yanase1, Yanase2}:
\begin{equation}
    {j}^{y}_{\rm BCD}(t) =\tilde{\chi}_0 q^x  \dot q^x,\quad {j}^{x}_{\rm BCD}(t) = -\tilde{\chi}_0 q^x \dot q^y,
    \label{eq:NHE}
\end{equation}
where $\tilde{\chi}_0 = \tilde{\chi}^{yxx} = -\tilde{\chi}^{xxy}$. In what follows, we will focus on two-dimensional materials with a mirror symmetry $\mathcal{M}_x:x\rightarrow-x$ (mirror along $y$),  the highest point group symmetry that permits a non-zero antisymmetric $\tilde{\chi}^{\gamma\alpha\beta}_- = e {\mathcal{D}}^\alpha\epsilon^{\beta \gamma}$. We term such currents ``BCD'' because in noncentrosymmetric and non-magnetic superconductors, it is directly controlled by the superconducting Berry curvature dipole $\boldsymbol{\mathcal{D}} = (\mathcal{D},0)$ (BCD)~\cite{sBCD_accepted} perpendicular to the mirror, where $\mathcal{D}$ has units of length. Here $\mathcal{D}$ is the Berry curvature dipole strength and $\tilde{\chi}_0= e\mathcal{D}$. In what follows, we will focus on this BCD dissipationless nonlinearity and its manifestation in a quantum circuit.

\vspace{2mm}
\textit{\textcolor{blue}{BCD nonlinear quantum circuit.}} The constitutive relations in Eq.~(\ref{eq:NHE}) describe an intrinsic bulk nonlinearity in noncentrosymmetric superconductors. As we now argue, these nonlinear currents can be used to produce a nonlinear quantum circuit without Josephson junctions and without magnetic fields. 
  
To see this, consider a superconductor of size $\boldsymbol{\ell} = [\ell_x,\ell_y]$ with uniform dynamical supercurrent flows. The condensate momentum $2\hbar \mathbf{q}$ is then related to the phase drop across the superconductor as (e.g., in the $x$ direction) $\phi_x(t)=2\int_0^{\ell_x}[q(t)]^x dx=\theta_2(t)\!-\!\theta_1(t)\!-\!\frac{2e}{\hbar}\!\int_0^{\ell_x} [A(t)]^x d x,$ where $\theta_1 (t)$ and $\theta_2 (t)$ denote superconducting condensate phases at two ends of the sample (at $x = 0$ and $x = \ell_x$) and $\mathbf{A} (t)$ is a uniform oscillating vector potential. As a result, for a superconductor with a uniform total supercurrent we write $\phi_x(t) = 2\ell_x q^x (t)$ and $\phi_y(t) =2 \ell_y q^y (t)$. These enable a re-expression of Eq.~(\ref{eq:NHE}) in terms of the circuit phase coordinates and their time derivatives. 

In particular, within a lumped circuit element approach, the BCD nonlinear currents can be naturally described via a third-order Lagrangian term
\begin{equation}
\mathcal{L}_{\rm BCD} (t) =  \hbar \tilde{\mathcal{D}} \phi_x (t) \dot{\phi}_x (t)\phi_y (t), \quad \tilde{\mathcal{D}} = \mathcal{D}/(8\ell_x),
\label{eq:main_deltaL}
\end{equation} 
where the Euler-Lagrange equations with a current read as $d/dt (\partial  \mathcal{L}/\partial \dot{\phi}_\alpha) - \partial \mathcal{L}/\partial \phi_\alpha = (\hbar/2e)I_\alpha$; substituting $\mathcal{L}_{\rm BCD}$ in Eq.~(\ref{eq:main_deltaL}) produces the nonlinear currents Eq.~(\ref{eq:NHE}). Notice that two orthogonal $x$- and $y$-directed phase drops couple in $ \mathcal{L}_{\rm BCD}$: this means that the nonlinear coupling enabled by $ \mathcal{L}_{\rm BCD}$ are active in multi-terminal superconducting devices where modes of different spatial structure coexist.

\begin{figure}[t]
\centering
\begin{minipage}[t]{0.98\linewidth}
\vspace{0.2em}
\centering
\includegraphics[width=\linewidth]{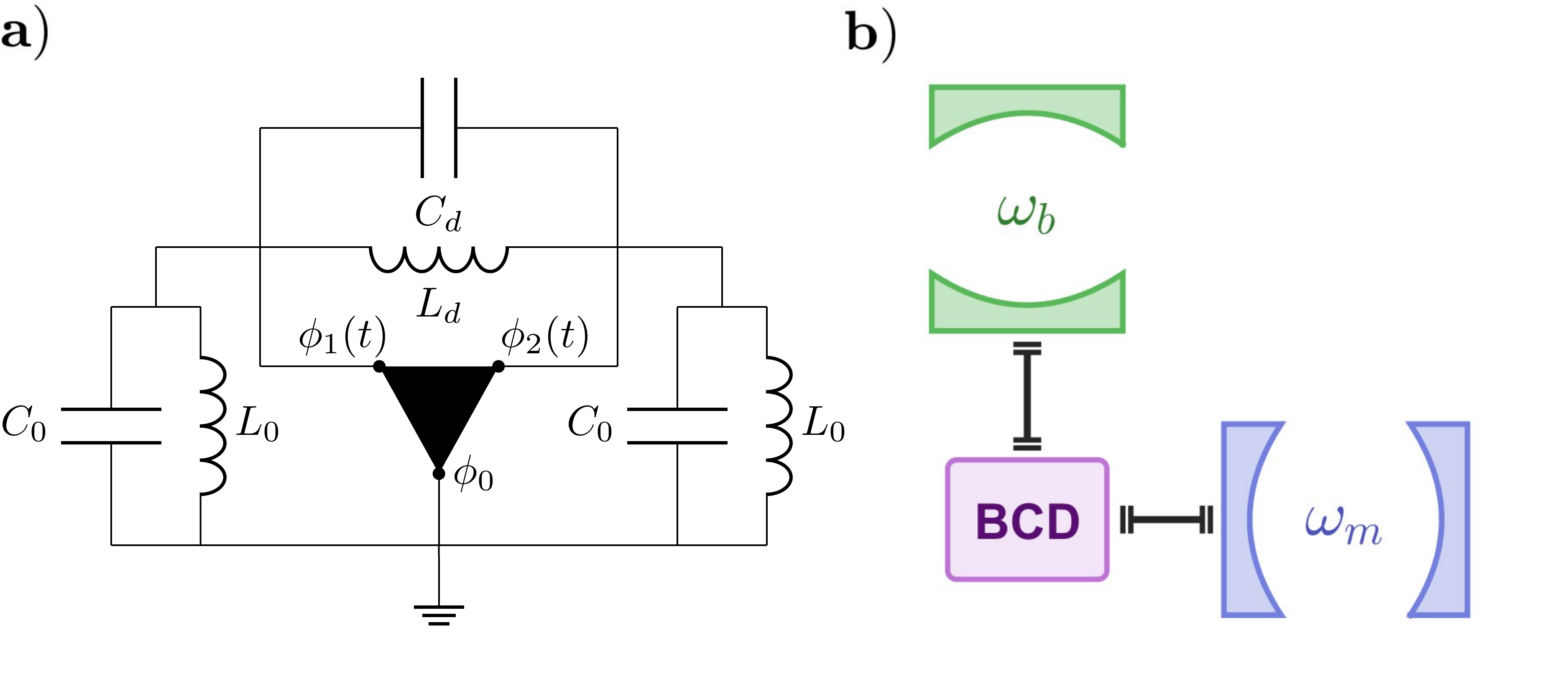}
\end{minipage}
\\
\begin{minipage}[t]{0.98\linewidth}
\vspace{0.2em}
\centering
\includegraphics[width=\linewidth]{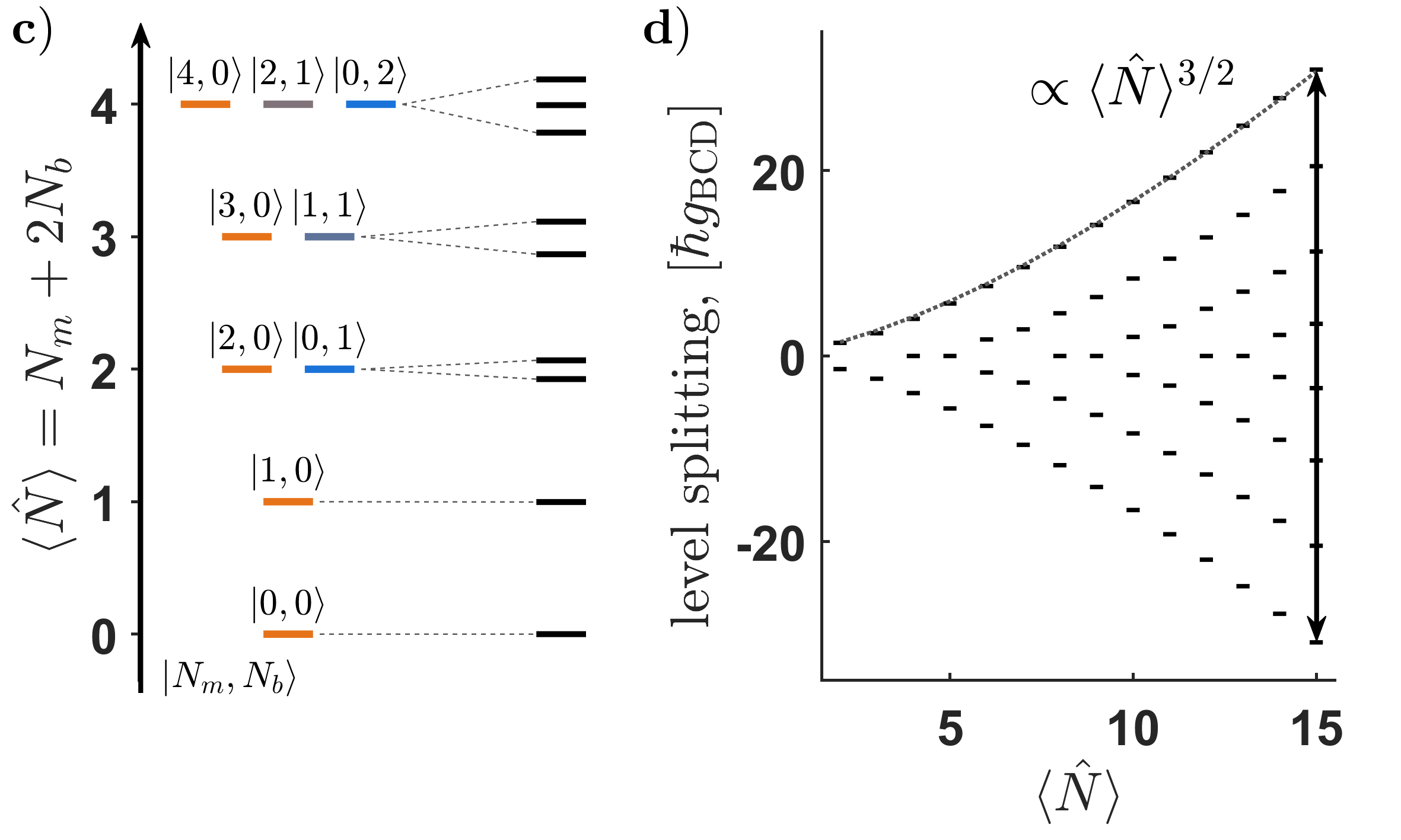}
\end{minipage}
\caption{\textbf{BCD induced anharmonicity and multiphoton vacuum level splitting.} (a) Minimal three-terminal circuit implementation of the BCD induced nonlinear mode mixing. (b) Illustration of the buffer and memory modes coupled by the BCD nonlinear quantum circuit element. (c) (left) Energy levels of uncoupled memory and buffer modes at an autoparametric resonance $\omega_b = 2\omega_m$ become mixed and (right) split due to the quantum three-wave mixing in Eq.~(\ref{eq:sBCD_quantized_hamiltonian}) that couple a buffer mode photon with two memory mode photons. The black levels were obtained from an exact diagonalization of $\hat H = \hat H_0 + \hat H_{\rm BCD}$ for $g_{\rm BCD}/\omega_m=0.05$. Here $N_m = \langle \hat{m}^\dagger \hat{m}\rangle$ and $N_b = \langle \hat{b}^\dagger \hat{b} \rangle $. (d) The splitting grows with photon number scaling as $\delta E^{(1)}/(\hbar g_{\rm BCD})\propto \langle \hat N  \rangle^{3/2}$ (dashed line); here solid black levels were obtained from diagonalizing $\hat{H}$ within each $N$-indexed degenerate sub-space.}\label{sBCDQFIG2}
\end{figure}

We illustrate this in a minimal three-terminal geometry shown in Fig.~\ref{sBCDQFIG1}(b) and Fig.~\ref{sBCDQFIG2}(a) where phase drops $\phi_\alpha$ accrue inductive energy $E^\alpha_{L} \phi_\alpha^2/2$ and  their conjugate momenta $n_\alpha$ accrue capacitive energy $2E^\alpha_{C} n_\alpha^2$. Using the BCD nonlinear superconductor element, and writing the phase drops across the superconductor in terms of the node phases $\phi_{0,1,2}$ so that $\phi_1= \phi_0+\phi_y-\phi_x/2$ and $\phi_2=\phi_0+\phi_y+\phi_x/2$, we obtain a nonlinear circuit Hamiltonian:  
\begin{equation}
   H\! = \! 2E^x_{C}(n_x \!-\! \tilde{\mathcal{D}}\phi_x \phi_y)^2 \!+ \frac{E^x_{L}}{2} \phi_x^2\!+\!2E^y_{C}n_y ^2 \!+ \frac{E^y_{L}}{2} \phi_y^2, 
\label{eq:sBCD_current}
\end{equation}
where $E_{C}^y = e^2/2C_0$, $E_C^{x}=2e^2/(C_0+2C_d)$, $E_L^y =(2/L_0)(\hbar/[2e])^2 + E^y_{L,{\rm SC}}$ and $E_L^x=(1/[2L_{0}]+1/L_d)(\hbar/[2e])^2 + E^x_{L,{\rm SC}}$ (see Fig.~\ref{sBCDQFIG2}(a)); here $E^{x,y}_{L,{\rm SC}}$ are the inductive energies of the superconductor along $x,y$ directions. This yields an anharmonic circuit Hamiltonian that couples the $x$ and $y$ modes. See Supplemental Material {\bf SM} for a detailed derivation of Eq.~(\ref{eq:sBCD_current}) and discussion of implementation in other circuit geometries.

We remark now on two important aspects of Eq.~(\ref{eq:sBCD_current}). First, at a fundamental level, the form of Eq.~(\ref{eq:sBCD_current}) can be viewed as a type of circuit-level ``minimal coupling'': $n_x \to n_x - \tilde{\mathcal{D}}\phi_x \phi_y$ where the nonlinear $\tilde{\mathcal{D}}\phi_x \phi_y$ produces nonlinear Hall currents. Notice that a minimal coupling that is quadratic in phase is the simplest replacement that produces anharmonicity in $H$: replacements linear or constant in the phases merely rotate the coordinates, leaving the circuit Hamiltonian quadratic. Second, expanding the square in Eq.~(\ref{eq:sBCD_current}) reveals a leading cubic nonlinear interaction proportional to $n_x\phi_x\phi_y$. The charge $n_x$ indicates the \textit{dynamical} origin of the BCD mixing. 

The circuit Hamiltonian in Eq.~(\ref{eq:sBCD_current}) can be quantized. To proceed, we promote $\phi_x\rightarrow\hat{\phi}_m$ and $\phi_y\rightarrow \hat{\phi}_b$ with their conjugate quantum operators satisfying $[\hat{\phi}_i,\hat{n}_j]=i\delta_{ij}$. We then write $\hat{\phi}_m =\varphi_{\rm zpf,m}(\hat m +\hat m^\dagger)$, $\hat n_m=i (\hat m^\dagger-\hat m)/(2\varphi_{\rm zpf,m})$, where $\hat m$ is the lowering operator for the mode with zero-point phase fluctuations $\varphi_{\rm zpf,m}=(E^x_{C}/E^x_{L})^{1/4}$ and frequency $\omega_m =2 \sqrt{E^x_{C}E^x_{L}}/\hbar$. Similarly we define $\hat b$ as the lowering operator for the mode with zero-point phase fluctuations $\varphi_{\rm zpf,b}=(E^y_{C}/E^y_{L})^{1/4}$ and frequency $\omega_b =2 \sqrt{E^y_{C}E^y_{L}}/\hbar$. These two modes define the harmonic part of the Hamiltonian: $\hat{H}_0 = \hbar \omega_b (\hat b^\dagger \hat b + 1/2) +\hbar \omega_m (\hat m^\dagger \hat{m} + 1/2) $. As a result, we obtain $\hat{H} = \hat{H_0}+ \hat{H}_{\rm BCD}$ with 
\begin{equation}
\hat{H}_{\rm BCD}=i\hbar g_{\rm BCD}\left[\hat m^2-\hat m^{\dagger 2}\right]\hspace{-0.5mm}(\hat b+\hat b^\dagger), 
\label{eq:sBCD_quantized_hamiltonian}
\end{equation}
where $g_{\rm BCD} = \mathcal{D} \omega_m \varphi_{{\rm zpf},b}\varphi^2_{{\rm zpf},m}/(8\ell_x)$ is the BCD induced nonlinear coupling between the two modes in the quantum limit. A higher order $\mathcal{O}(\phi^4)$ term can be generated from~\eqref{eq:sBCD_current}, however, these are not only higher order in $\varphi_{\rm zpf}$ but also higher order in $\mathcal{D}$, see also discussion in {\bf SM}. 

Eq.~(\ref{eq:sBCD_quantized_hamiltonian}) is the central result of this paper: a time-reversal-symmetric noncentrosymmetric superconductor realizes a three-wave-mixing quantum circuit nonlinearity $\hat{H}_{\rm BCD}$ through the superconducting Berry curvature dipole. Importantly, this three-wave mixing BCD quantum nonlinearity arises without any junctions. Instead, it emerges from the noncentrosymmetric order in an otherwise uniform superconductor: inversion breaking order in the material enables junction free nonlinearity. $\hat{H}_{\rm BCD}$ demonstrates how BCD nonlinear currents, that are often understood as {\it semiclassical} responses~\cite{sBCD_accepted,SodemannFu}, persist as a {\it quantum nonlinearity} at the single quantum mode level. 

As a demonstration of this nonlinearity, we simulate the static energy spectrum for $\hat{H} = \hat{H}_0 + \hat{H}_{\rm BCD}$ choosing $\omega_b = 2\omega_m$. When $g_{\rm BCD} = 0 $, the spectrum exhibits a set of equally spaced harmonic oscillator levels with degeneracies between states with the same total quantum number $\hat N = \hat m^\dagger \hat m + 2 \hat b^\dagger \hat b$. Nonzero $g_{\rm BCD}$ splits those degeneracies, as shown visually in Fig.~\ref{sBCDQFIG2}(c). Importantly, the magnitude of the splitting grows with the occupation of the levels in a nonlinear fashion with $\langle \hat N  \rangle$ (Fig.~\ref{sBCDQFIG2}(d)). This characteristic change in spectrum can be used as a means of directly extracting $\mathcal{D}$ in superconductors by employing a high-coherence microwave circuit and measuring the BCD-induced resonance splittings as in Fig.~\ref{sBCDQFIG2}(d). Recently, microwave resonance techniques have been used to extract the quantum metric by measuring the stiffness in superconductors in moiré material platforms \cite{Tanaka2025}. Fig.~\ref{sBCDQFIG2}(d) demonstrates these techniques can be applied to other quantum geometric characteristics such as the BCD.

\begin{figure}[t]
    \centering
    \includegraphics[width=0.5\textwidth]{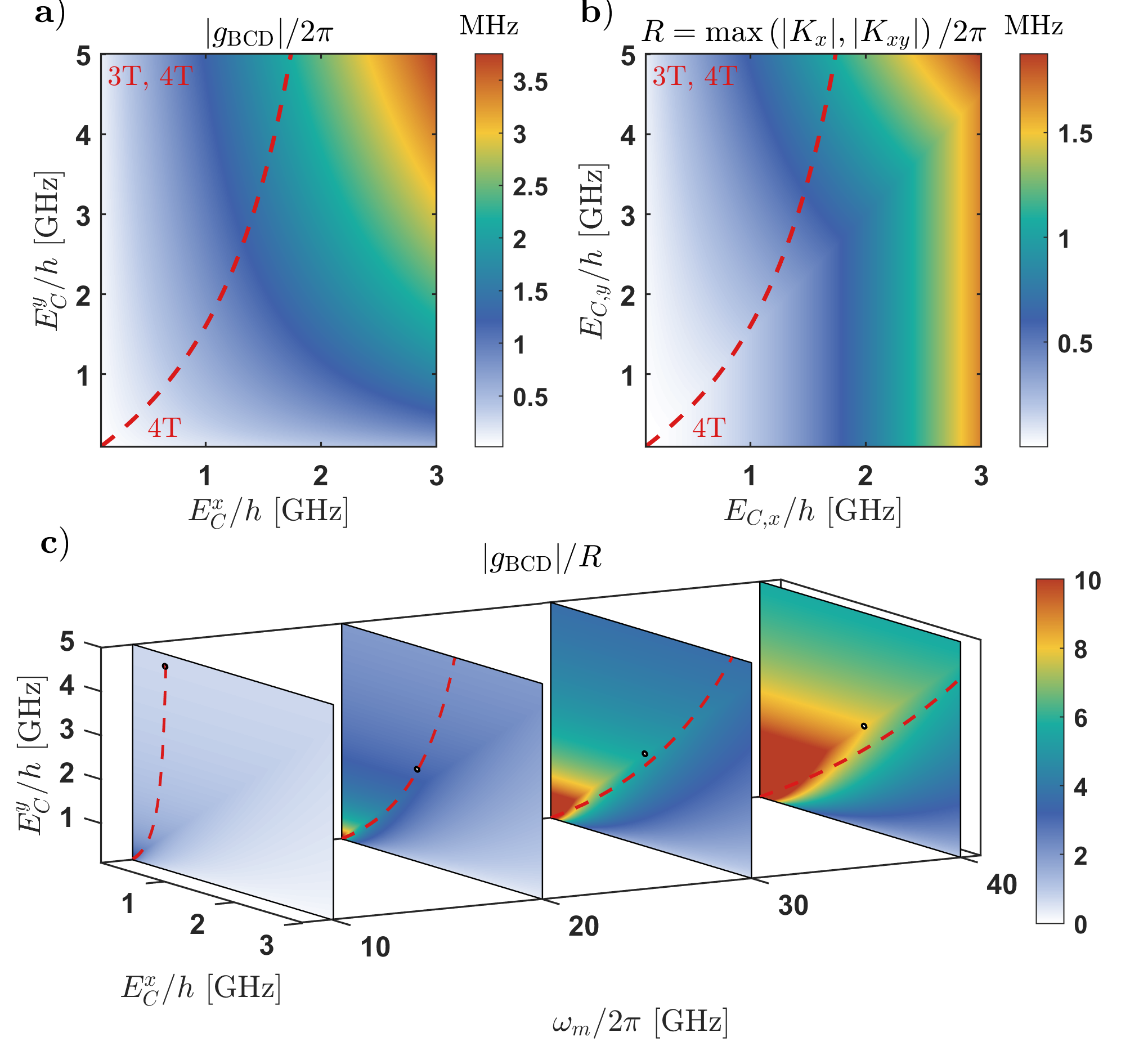}
    \caption{\textbf{BCD quantum nonlinearity in TBG/hBN superconducting platform.} (a) Sizeable BCD quantum three-wave mixing strength of several MHz can be achieved in a twisted bilayer graphene/hexagonal Boron Nitride superconducting device using $\hat{H} =  \hat{H}_0 + \hat{H}_{\rm BCD}$ for the three terminal circuit shown in Fig.~\ref{sBCDQFIG2}(a). $g_{\rm BCD}$ can be tuned by external capacitance $E^x_{C}$ and $E^y_{C}$. Here, and below, the red dashed line is a separatrix: $E^x_{C}$,$E^y_{C}$ values above the red dashed line can be achieved by both three terminal (3T) and four-terminal (4T) devices, see {\bf SM}. 4T devices can achieve all parameter regimes shown, see also discussion of 4T nonlinear BCD quantum Hamiltonian in {\bf SM}. (b) Estimate of intrinsic nonlinear Kerr effects (self Kerr and cross Kerr) effects in TBG/hBN superconducting device (see Eq.~(\ref{APKERR}) as a function of $E_{C,x}$ and $E_{C,y}$. For panels (a) and (b) we used $\omega_m/(2\pi)=20$ GHz. (c) Importantly for the parameter accessible regimes shown, BCD mixing can be up to an order of magnitude stronger then Kerr nonlinearities at various $\omega_m$ frequencies. On panel (c), dots represent points with $g_{\rm BCD}/2\pi = 1$ MHz and $|g_{\rm BCD}|/R= [0.7,3,5.5,8.1]$ corresponding to $\omega_m/(2\pi)=[10,20,30,40]$ GHz respectively. The range of $E_L/h$ on panels (a) and (b) is $[30,5000]$ GHz. The region with large $E_L$ (equivalently smaller $E_C$) matches the region of large values of $|g_{\rm sBCD}|/R$ shown in panel (c). In all plots we have tuned to the auto-parametric regime $\omega_b = 2 \omega_m$, used device size $\ell_x=\ell_y = 2\mu$m, and estimated tBG/hBN Berry curvature dipole as $\mathcal{D}=20$nm, intrinsic Kerr strengths $\mathcal{K}_{xxxx}= 3\times 10^5$ GHz nm$^{-2}$, $\mathcal{K}_{xxyy}= 10^5$ GHz nm$^{-2}$ and set $E^x_{L,{\rm SC}}/h=E^y_{L,{\rm SC}}/h=50$ GHz. A detailed discussion of our numerical simulation of the superconducting quantum nonlinearities in TBG/hBN can be found in {\bf SM}. }\label{sBCDQFIG3}
\end{figure}

\vspace{2mm}
\textit{\textcolor{blue}{BCD bosonic qubit.}} Even as the anharmonic levels of a BCD device allow isolation of a distinct pair of levels that can be used as a qubit, the form of $\hat{H}_{\rm BCD}$ make it particularly suited for a different type of qubit implementation: driven-dissipative stabilization of bosonic qubits. Bosonic codes encode quantum information into superpositions of states in the large Hilbert space spanned by a quantum harmonic oscillator \cite{SciPostPhysLectNotes.72}. 

Here, as an example, we inspect Schr\"odinger cat qubits \cite{MirrahimiIOP_2014,SciPostPhysLectNotes.72,Grimm2020,MirrahimiMixedCat} stabilized through a three-wave mixing-based protocol developed by Marquet, et al.~\cite{autocat}. The buffer mode in Eq.~(\ref{eq:sBCD_quantized_hamiltonian}) is driven as $\hat{H}_1 = \hat{H}_0 + \hat{H}_{\rm BCD} + F_b e^{-i\Omega_b t}\hat b^\dagger+F_b^\ast e^{i\Omega_b t}\hat b$ where $F_b$ and $\Omega_b$ are the drive amplitude and drive frequency respectively. Choosing $\Omega_b = \omega_b$ and applying the rotating wave unitary transformation $U(t) = {\rm exp}[- i \omega_m t \hat{m}^\dagger \hat{m} - i \omega_b t \hat{b}^\dagger \hat{b}]$ (applied as $\tilde H = U^\dagger\hat{H}_1 U - i\hbar U^\dagger \dot{U}$) produces
\begin{equation}
    \tilde{H} = \hat{b}^\dagger [\hbar g_{\rm BCD} \hat{m}^2 + F_b] + \hat{b} [\hbar g_{\rm BCD} \hat{m}^{\dagger 2} + F_b^*], 
    \label{eq:RWA}
\end{equation}
where we have chosen an `auto-parametric' $m,b$ level configuration $\omega_b = 2\omega_m$ (same as in Fig.~\ref{sBCDQFIG2}(c)). By parameterizing $F_b = -\hbar g_{\rm BCD} \alpha^2$, we see the relationship between the drive and the two nearly-degenerate coherent states stabilized in the memory mode $|\pm\alpha\rangle$ are zero-modes of $\tilde{H}$ \cite{autocat,SciPostPhysLectNotes.72}. Such states are attractive because they can be exponentially disjoint $\langle +\alpha|(-\alpha) \rangle = {\rm exp}\left(-2 |\alpha|^2\right)$ for large $\alpha$, suppressing logical bit flips that typically occur due to photon loss. 

The structure of Eq.~(\ref{eq:RWA}) is a two-photon driven-dissipative stabilization strategy \cite{autocat} to stabilize the coherent states. We give a brief outline of the concept. For a lossy buffer, the density matrix evolves according to $\hbar\dot{\rho} = -i[\tilde{H}, \rho] + \hbar\kappa_b \mathcal{L}(\hat{b})\rho$~\cite{SciPostPhysLectNotes.72, Azouit_2017,carmichael2009statistical} where $\rho$ is the full density matrix, $\kappa_b$ is the loss rate of the buffer, and $\hat{\mathcal{L}}$ is the Lindbladian dissipator. Taking $\kappa_b \gg g_{\rm BCD}|\alpha|$ and adiabatically eliminating the buffer mode dynamics, the memory mode density matrix evolves as $\dot{\rho}_m =  \kappa_2 \hat{{\mathcal{L}}}(\hat{m}^2 - \alpha^2) \rho_m$ where $\kappa_2 = 4 g_{\rm BCD}^2/\kappa_b$ and  $\hat \rho_m \equiv {\rm Tr}_{b}[\hat\rho]$~\cite{SciPostPhysLectNotes.72, Azouit_2017,carmichael2009statistical}. This ensures the zero-mode coherent states of $\tilde{H}$ are stationary states as well as {\it stabilized}, with $\kappa_2$ the stabilization rate, i.e., the continuous pushing of the system back to the coherent state (cat) manifold~\cite{SciPostPhysLectNotes.72}.

\vspace{2mm}
\textit{\textcolor{blue}{BCD quantum three-wave mixing in TBG superconducting platform.}} We now illustrate $g_{\rm BCD}$ in a two-dimensional noncentrosymmetric superconducting platform. We focus on strained twisted bilayer graphene (TBG) on hexagonal Boron Nitride (hBN) heterostructure that can have a large Berry curvature dipole \cite{PacoTBGBCD,KTLAW_TBG_BCD}; here hBN breaks centrosymmetry and strain breaks rotational symmetry to activate BCD. To model superconducting BCD~\cite{sBCD_accepted} in such a heterostructure, we use a simple model where superconducting order is induced by proximity effect to a superconductor layer. We find large superconducting BCD values $\mathcal{D}$ of order several tens of nanometers, see {\bf SM} for full calculation details. That said, we note that in real systems,e superconductivity appears intrinsically without the proximal layer~\cite{Tanaka2025,Cao2018,Yankowitz}.

We now turn to estimating the BCD quantum three-wave mixing strength in the simple three-terminal device structure Fig. \ref{sBCDQFIG3}(a). Recalling Eq.~(\ref{eq:sBCD_quantized_hamiltonian}) and the form of the zero-point fluctuations, we find
\begin{equation}
\hbar g_{\rm BCD} = \frac{\mathcal{D}}{4 \ell_x}  \sqrt{ \frac{(E_{C}^{x})^2 E_{C}^y}{\hbar \omega_m}}
\end{equation}
that produces $g_{\rm BCD}$ of order several MHz, as shown in Fig. \ref{sBCDQFIG3}(b), for example for parameter choice $\ell_x = 2\mu$m, $\mathcal{D}=20$ nm, $\omega_m/(2\pi) = 20$ GHz, $E_C^x/h = 1.7$ GHz, $E_C^y/h = 4.5$ GHz we find $g_{\rm sBCD}/(2\pi)=2$ GHz. This $g_{\rm BCD}$ mixing strength allows sizeable stabilization rates $\kappa_2$ of order several MHz necessary for driven-dissipative stabilized BCD cat qubit operation (here we took an illustrative buffer loss of $\kappa_b = 20 \, {\rm MHz}$). We remark that $g_{\rm BCD}$ scales with both $\ell^{-1}$ and $\omega_m^{-1/2}$. This means that $g_{\rm BCD}$ can be made larger by choosing smaller device sizes. Similarly, the scaling as $\omega_m^{-1/2}$ (stemming from the dynamical origin of the BCD nonlinear mixing) indicates that $g_{\rm BCD}$ can remain strong at higher frequencies (see next paragraph). In contrast, three photon interactions in magnetic flux-biased Josephson junctions scale as $\omega_m^{-3/2}$, rapidly diminishing at high frequency. 

In driven-dissipative stabilized bosonic qubits, it is essential to include the lowest-order  four-photon `Kerr' interactions that shift the energies of Fock states at high energy, e.g., $\hat{H}_{\rm Kerr}^{(m)} = \hbar K_x\hat m^{\dagger 2} \hat m^2 + \hbar K_{xy}\hat m^\dagger\hat m\hat b^\dagger\hat b$~\cite{Grimm2020,MirrahimiMixedCat}. Here, we must account for the quantum material origin of such Kerr nonlinearities. Writing $j_{\rm Kerr}=({8e}/{3\hbar}){\mathcal{K}}_{\gamma\alpha\beta\delta} q_\alpha q_\beta q_\delta$ we find 
\begin{equation}\label{APKERR}
     \hbar K_{x}\! = \!\frac{\ell_y\mathcal{K}_{xxxx}}{ \ell_x^3} \frac{E^2_{C,x}}{\hbar ^2\omega_m^2}, \quad \hbar K_{xy} \!= \!\frac{\mathcal{K}_{xxyy}}{ \ell_x\ell_y}\frac{2E_{C,x}E_{C,y}}{\hbar ^2\omega_m^2}
\end{equation}
where $\mathcal{K}_{xxxx}$ and $\mathcal{K}_{xxyy}$ are intrinsic material determined Kerr strengths, see {\bf SM}. Simulating $\mathcal{K}_{xxxx}$ and $\mathcal{K}_{xxyy}$ for the TBG/hBN superconducting heterostructure, we find the maximum Kerr strengths in the three-terminal device are of order 1 MHz (Fig.~\ref{sBCDQFIG3}(b)). Critically, $g_{\rm BCD}$ and $K_x, K_{xy}$ have distinct operating frequency and dimension scaling: the intrinsic Kerr strengths scale as $(\omega_m \ell)^{-2}$. As a result, the ratio $g_{\rm BCD}/{\rm max}[K_x, K_{xy}]$ can be made large by taking advantage of the scaling $\sim\omega_m^{3/2} \ell$ (see Fig.~\ref{sBCDQFIG3}(c)). Indeed, a window of large $g_{\rm BCD}/{\rm max}[K_x, K_{xy}]$ ratio rapidly expands with increasing operating frequency $\omega_m$, suggesting higher frequency superconducting circuits as a particularly promising regime for BCD quantum circuit elements where Kerr interactions are undesirable. 

\vspace{2mm}
\textit{\textcolor{blue}{Discussion.}} It is useful to contrast the BCD quantum nonlinearity with Josephson junction nonlinearities employed in quantum circuits. To do so, we first compare to a three-terminal Josephson junction device~\cite{autocat}, where a similar three-wave mixing can be realized with mixing strength $\hbar g_{\scalebox{0.6}{$\bcancel{\mathcal{T}}$}}  \propto  \varphi_{{\rm zpf},b}\varphi^2_{{\rm zpf},m} \sin \left( \bar \phi_J(\phi_\mathrm{ext}) \right)$, where $\phi_\mathrm{ext}$ is a magnetic flux penetrating a loop [see Fig.~\ref{sBCDQFIG1}(a)], and $\bar \phi_J$ is an antisymmetric function of $\phi_\mathrm{ext}$. This nonlinearity relies on breaking TRS via the external magnetic flux $ \phi_{\rm ext}$ and vanishes when the flux is zero. In contrast, $\hat{H}_{\rm BCD}$ arises even under TRS conditions in the absence of applied magnetic flux. Central to this nonlinearity are the BCD-enabled unconventional flux-charge type coupling in Eq.~(\ref{eq:main_deltaL}) that allows three photon interactions terms with TRS. 

In a second comparison, we consider field-effect tunable Josephson elements (sometimes called JoFETs), which have recently been used to demonstrate flux-charge coupling in a quantum circuit~\cite{joyez_single_1995,leroux_nonreciprocal_2022,isakov_observation_2026}. The lowest-order nonlinearity from such flux-charge coupling is of the form $\dot{\phi}_x \phi_y^2$ and, like the BCD, can persist without magnetic field. Although the circuit-level coupling is related to our BCD-based Lagrangian Eq.~(\ref{eq:main_deltaL}) via a full time-derivative (see {\bf SM}), the experimental differences are meaningful. Electric field tuning of Josephson energy in JoFETs requires highly concentrated electric fields in the field-effect gate electrodes which may produce significant dielectric losses~\cite{hung_probing_2022}, and electric field tunability also implies sensitivity to charge noise (especially when preserving $\mathcal{T}$, see~\cite{isakov_observation_2026}). Field-effect tunable JoFETs also carry concerns about other materials-related sources of energy relaxation~\cite{strickland_characterizing_2024,sun_junction-intrinsic_2026,lu_fabrication-free_2026}. Our proposed BCD nonlinear quantum circuit element, in contrast, relies instead on the intrinsic nonlinear dynamics embedded in a quantum material, which may offer new avenues. 

Our work demonstrates how nonlinear dynamics in noncentrosymmetric quantum materials can augment superconducting quantum circuits. In particular, the BCD nonlinear Hall effect persists in the quantum limit to mediate quantum three-wave mixing even without magnetism. This provides a concrete example of using quantum materials as a physical resource for quantum information processing and devices. While we have focused on the BCD nonlinearity, we anticipate a range of other types of nonlinearities enabled by material quantum geometry can similarly be exploited. For example, when time-reversal symmetry is broken, another Hall-type  nonlinearity can be activated: Berry connection polarizability \cite{PhysRevResearch.2.043081,PhysRevLett.127.277202,PhysRevLett.127.277201}. We remark further that multiterminal Josephson devices with open internal regions to enable the Hall current could be useful~\cite{zhang_nonlocal_2026} -- if the normal material is appropriately noncentrosymmetric, then a Josephson version of the BCD circuit element may also be possible; engineering such configurations are a promising direction for future work. These examples underscore the richness of quantum materials-enabled dynamics for use in superconducting circuit quantum electrodynamics.

\textit{Acknowledgements}. V.F. acknowledges helpful discussions with Baptiste Royer and Benjamin Huard. J.C.W.S. acknowledges useful conversations with Shicong Wang. This research is supported by the Singapore Ministry of Education (MOE) under its Academic Research Fund Tier 2 grant (MOE-T2EP50222-0011), Academic Research Fund Tier 2 grant MOE-T2EP50225-0008 and Academic Research Fund Tier 3 grant MOE-MOET32023-0003 ``Quantum Geometric Advantage''.

\bibliography{BCDqubit}
\clearpage
\onecolumngrid
\clearpage
\appendix
\onecolumngrid
\setcounter{equation}{0}
\setcounter{page}{1}
\setcounter{figure}{0}
\renewcommand{\theequation}{\thesection-\arabic{equation}}
\renewcommand{\thefigure}{\thesection-\arabic{figure}}
\renewcommand{\thetable}{\thesection-\Roman{table}}
\titleformat{\section}[block]{\normalfont\normalsize\bfseries\centering}{Section \Alph{section}:}{0.3em}{}
\section*{\large Supplemental Material for ``Berry curvature dipole circuit quantum electrodynamics''}

\section{Lagrangian and Hamiltonian of the three and four terminal device with BCD mixing element}
In this Supplementary Material section, we provide an extended discussion of the effective circuit Lagrangian generated by the BCD mixing.

As a simple exercise (and complementary to that discussed in the main text), we first recount how a BCD second order current can be expressed in terms of phases. We begin superconducting BCD response~\cite{sBCD_accepted,Yanase1, Yanase2}, where in the low-frequency limit $\hbar\omega\ll\Delta_{\rm SC}$ the nonlinear Hall current in a superconductor can be generally written as
\begin{equation}
    {j}^{\gamma}_{\rm BCD}(\omega_2)=i\frac{e^3}{\hbar^2}\mathcal{D}^{\alpha\beta\gamma} \int  \frac{d\omega_1}{2\pi}E_{\alpha}(\omega_1)E_\beta(\omega_2-\omega_1)\frac{1}{\omega_1}.
\end{equation}
with $\mathcal{D}^{\alpha\beta\gamma} = \sum_a f_{\rm FD} [\mathcal{E}_{a,\mathbf{k}}(\mathbf{q})] \partial_\mathbf{q}^\alpha \bra{\partial_\mathbf{q}^\beta \Psi_a(\mathbf{k,q})} \partial_\mathbf{q}^\gamma \Psi_a (\mathbf{k,q})\rangle - (\beta\leftrightarrow\gamma)$ (see section C for more details). Here we use the Fourier convention $f(t)=\int d\omega_1\, e^{-i\omega_1 t}f(\omega_1)/(2\pi)$.

We next convert dynamical electric fields into voltages via $E_\alpha(t)=V_\alpha(t)/\ell_\alpha$, and later to phases using the Josephson relation $V^\alpha=-(\hbar/2e)\dot\phi_\alpha$. Such follows from the gauge-invariant phase difference across the superconducting device, $\phi_x(t)=\phi_2(t)-\phi_1(t)=\theta_2(t)-\theta_1(t)-(2e/\hbar)\int_l \mathbf A\cdot d\mathbf l$. Here $\theta_1$ and $\theta_2$ denote microscopic superconducting condensate phase while $\phi_1$ and $\phi_2$ are circuit node phases in the convention used in Fig.~\ref{sBCDQFIG1}(b), so that the electromagnetic line-integral contribution is absorbed into the gauge-invariant circuit coordinate $\phi_x$. Taking the time derivative gives the Josephson voltage relation we find $\dot\phi_\alpha(t)=(2e/\hbar)\int_l \mathbf E\cdot d\mathbf l=(2e/\hbar)V_\alpha$. By transforming the nonlinear response back to the time domain we find
\begin{align*} 
    {j}^{\gamma}_{\rm BCD}(t)&=\frac{ie^3}{\hbar^2 \ell_\alpha \ell_\beta}\mathcal{D}^{\alpha\beta\gamma}\int\frac{d\omega_2}{2\pi}e^{-i\omega_2 t}\int \frac{d\omega_1}{2\pi}V_\alpha(\omega_1)V_\beta(\omega_2-\omega_1) \frac{1}{\omega_1}\\&=-\frac{ie}{4 \ell_\alpha \ell_\beta}\mathcal{D}^{\alpha\beta\gamma} \int \frac{d\omega_2}{2\pi} e^{-i\omega_2 t}\int\frac{d\omega_1}{2\pi}(\omega_2-\omega_1)\phi_{\alpha}(\omega_1)\phi_\beta(\omega_2-\omega_1)\\&=-\frac{ie}{4 \ell_\alpha \ell_\beta}\mathcal{D}^{\alpha\beta\gamma} \int \frac{d\omega_1}{2\pi}\phi_{\alpha}(\omega_1)\int \frac{d\omega_2}{2\pi} e^{-i(\omega_2-\omega_1+\omega_1) t}(\omega_2-\omega_1)\phi_\beta(\omega_2-\omega_1)\\&=-\frac{ie}{4 \ell_\alpha \ell_\beta}\mathcal{D}^{\alpha\beta\gamma}\int \frac{d\omega_1}{2\pi}e^{-i\omega_1 t}\phi_{\alpha}(\omega_1)\int \frac{d\bar\omega}{2\pi} e^{-i\bar\omega t}\bar\omega\phi_\beta(\bar\omega)=\frac{e}{4 \ell_\alpha \ell_\beta}\mathcal{D}^{\alpha\beta\gamma}\phi_\alpha(t)\dot\phi_\beta(t),
\end{align*}
For the mirror symmetric case considered in the main text there are only two nonzero BCD components $\mathcal{D}^{xxy}=-\mathcal{D}^{xyx}=\mathcal{D}$ thus for $I_x = j_x \ell_y,I_y = j_y \ell_x$ the total currents lowing two directions are therefore
\begin{equation}
    {I}^{y}_{\rm BCD}(t) = \frac{e}{4 \ell_x }\mathcal{D} \phi_x(t)\dot\phi_x(t),\qquad {I}^{x}_{\rm BCD}(t) = -\frac{e}{4 \ell_x}\mathcal{D} \phi_x(t)\dot\phi_y(t).\label{s2}
\end{equation}

Alternatively, Eq.~(\ref{s2}) can be directly obtained from Eq.~(\ref{eq:NHE}) of the main text. To do so, directly substitute the $\phi_\alpha(t) = \phi_x(t)=2\int_0^{\ell_\alpha}[q(t)]^\alpha dr^\alpha= 2\ell_\alpha q^\alpha(t)$ into Eq.~(\ref{eq:NHE}) of the main text. 

We now turn to discuss the Lagrangian. For uniform phase gradients, we write the Euler--Lagrange equation in the presence of a current source and find the minimal correction to Lagrangian that includes the BCD nonlinearity as
\begin{equation}
    \frac{d}{dt} \frac{\partial \delta\mathcal{L}}{\partial \dot q^\alpha} -\frac{\partial \delta\mathcal{L} }{\partial q^\alpha} =\frac{\hbar \ell_x\ell_y}{e } j^\alpha(t); \quad \frac{d}{dt} \frac{\partial \delta\mathcal{L} }{\partial \dot \phi_\alpha} -\frac{\partial \delta\mathcal{L} }{\partial \phi_\alpha} =\frac{\hbar}{2e} I^\alpha(t) \quad \Rightarrow \quad \delta\mathcal{L} _{\rm BCD}= \frac{\hbar}{8  \ell_x}\mathcal{D}\phi_x(t)\dot\phi_x(t) \phi_y(t).\label{s3}
\end{equation}
The argument for the above introduction of the current drive is to define the current source as a flux ($\phi_{\rm source}$) and large inductor ($L_{\rm source}$) connected in parallel to the circuit \cite{Vool_2017} with the full Lagrangian $\mathcal{L}_{\rm full}=\mathcal{L}_{\rm circuit}-(\hbar/[2e])^2(\phi-\phi_{\rm source})^2/(2L_{\rm source})$, where $\phi$ is the phase drop on the circuit. Then Lagrange equation $d(\partial \mathcal{L}_{\rm full}/\partial \dot \phi)/dt -\partial \mathcal{L}_{\rm full}/\partial \phi =0$ is equivalent to $d(\partial \mathcal{L}_{\rm circuit}/\partial \dot \phi)/dt -\partial \mathcal{L}_{\rm circuit}/\partial \phi =(\hbar/[2e])^2(\phi_{\rm source}-\phi)/L_{\rm source}$ that for large inductance $\phi/L_{\rm source}\rightarrow0$ with $(\hbar/[2e])\phi_{\rm source}/L_{\rm source}\rightarrow  I_{\rm source}$ plays the role of the current source. 

Generally, other forms of the Lagrangian are also allowed. For example, we can employ the fact that Lagrangians are defined up to the full time derivative allowing us to introduce a class of Lagrangians parametrized by $\zeta$ and written as $\delta\mathcal{L}^\zeta_{\rm BCD}=\hbar g_{\rm BCD}[(1-\zeta)\dot{\phi}_x\phi_x\phi_y -(\zeta/2)\phi_x^2 \dot \phi_y]$. Notice $\delta\mathcal{L}^{\zeta=1}_{\rm BCD}-\delta\mathcal{L}^{\zeta=0}_{\rm BCD}$ is a full time derivative, $\zeta$ does not affect equations of motion and currents. Lastly, we note that in Eq.~(\ref{s2}) and Eq.~(\ref{s3}) we have focused on the BCD nonlinearity in a system with mirror symmetry $\mathcal{M}_x$ perpendicular to $x$ axis. In such systems, the BCD nonlinearity is described by $(\mathcal{D},0)$. However, in systems which have lowered symmetries where mirror has been broken, Berry dipole will have two components $(\mathcal{D}_x,\mathcal{D}_y)$ allowing $\delta\mathcal{L}_{\rm BCD}\sim \chi^{\alpha\beta\gamma}_{\rm BCD}\dot{q}^\alpha q^\beta q^\gamma$. Nevertheless, the same procedure outlined above can still be used to obtain the effective Langrangian. In what follows, we focus on the simple mirror symmetric case. 

\begin{figure}[t]
\centering
\begin{minipage}[t]{0.49\linewidth}
{\raggedright \textbf{(a)}\par}
\vspace{0.2em}
\centering
\includegraphics[width=0.7\linewidth]{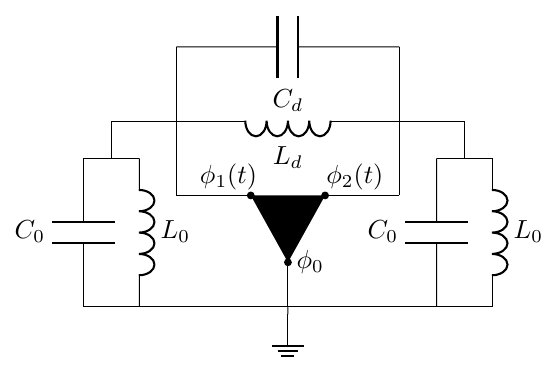}
\end{minipage}
\hfill
\begin{minipage}[t]{0.49\linewidth}
{\raggedright \textbf{(b)}\par}
\vspace{0.2em}
\centering
\includegraphics[width=0.7\linewidth]{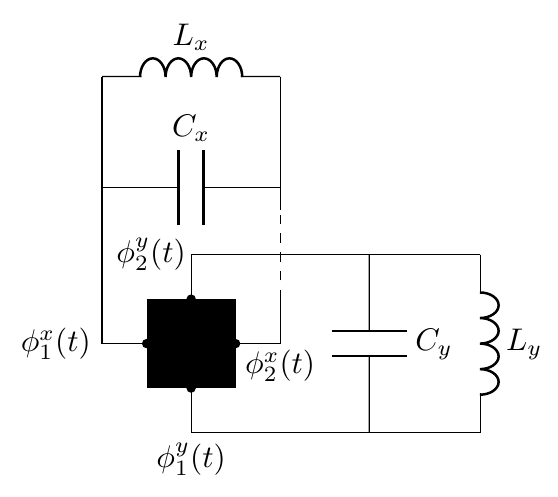}
\end{minipage}
\caption{\textbf{Schematic of BCD-based quantum three-wave mixer in three-terminal and four-terminal circuits.}
Schematic illustration of a) three terminal b) four terminal circuit topology with BCD mixing element. In three terminal embedding we write $\phi_1(t) = \phi_0 + \phi_y(t) - \phi_x(t)/2, \phi_2(t) = \phi_0 + \phi_y(t) + \phi_x(t)/2$ while in the four terminal device $\phi_x(t)=\phi^x_2(t)-\phi^x_1(t),\phi_y(t)=\phi^y_2(t)-\phi^y_1(t)$.}\label{fig:deviceSM}
\end{figure}

\subsubsection{Three terminal device as shown in Fig. \ref{fig:deviceSM}(a)} 
In this section we demonstrate that the Hamiltonian from Eq.~(\ref{eq:sBCD_current}) of the main text is consistent with the current density response in Eq.~(\ref{eq:NHE}) of the main text. We first consider the three-terminal geometry shown in Fig. (\ref{fig:deviceSM})(a) with horizontal and vertical directions to be identified as $x$ and $y$ respectively. Notice, in addition to external circuit inductance, the superconductor itself has finite intrinsic $E^\alpha_{\rm SC}$ inductive energy. We employ the following phase conventions: $\phi_0$ is ground phase, in what follows we will set it to zero without of loss of generality, $\phi_1 = \phi_y -\phi_x/2,\phi_2 = \phi_y +\phi_x/2$ to find the Lagrangian as: 
\begin{equation}
\begin{aligned}
    \mathcal{L}&=\frac{\hbar^2 \dot \phi_1^2 }{8 E_{C_0}}+ \frac{\hbar^2 (\dot \phi_1-\dot \phi_2)^2 }{8 E_{C_d}} + \frac{\hbar^2 \dot \phi_2^2 }{8 E_{C_0}}-\frac{E_{L_0}}{2} \phi_1^2-\frac{E_{L_0}}{2} \phi_2^2 - \frac{E_{L,d}}{2} (\phi_2 - \phi_1)^2- U_{\rm SC}\\&=\frac{\hbar^2 \dot \phi_x^2 }{8 E^x_C}+  \frac{\hbar^2 \dot \phi_y^2 }{8 E^y_C}-\frac{E^x_L}{2} \phi_x^2-\frac{E^y_L}{2} \phi_y^2 +\hbar g_{\rm BCD}((1-\zeta) \phi_x \dot\phi_x\phi_y -(\zeta/2)\phi_x^2\dot\phi_y)- U_{\rm Kerr}
\end{aligned}
\end{equation}
where $E_{C}^y = e^2/[2C_0]$, $E_C^{x}=2e^2/(C_0+2C_d)$, $E_L^y =(2/L_0)(\hbar/[2e])^2 + E^y_{L,{\rm SC}}$ and $E_L^x=(1/[2L_{0}]+1/L_d)(\hbar/[2e])^2 + E^x_{L,{\rm SC}}$ (see Fig.~\ref{sBCDQFIG2}(a) of the main text). For a superconductor with mirror symmetry $\mathcal{M}_x$ we have:
\begin{align}
    &U_{\rm SC} =- \frac{1}{2} E^x_{L,SC}\phi_x^2- \frac{1}{2} E^y_{L,SC}\phi_y^2 +\hbar g_{\rm BCD}((1-\zeta) \phi_x \dot\phi_x\phi_y -(\zeta/2)\phi_x^2\dot\phi_y)\!+ U_{\rm Kerr},\\
    &U_{\rm Kerr} =  \frac{\ell_x\ell_y}{4!}\!\left[\frac{\mathcal{K}_{xxxx}}{ \ell_x^4}\phi_x^4+\frac{6\mathcal{K}_{xxyy}}{ \ell_x^2 \ell_y^2}\phi_x^2\phi_y^2+\frac{\mathcal{K}_{yyyy}}{ \ell_y^4}\phi_y^4\right].
\end{align}
The intrinsic Kerr effect here is included here for completeness; a full discussion of the Kerr effect and how it emerges from supercurrent dynamics is presented in the next section. Note mirror symmetry $\mathcal{M}_x$ enforces $\mathcal{K}_{xyyy}=\mathcal{K}_{yxyy}=\mathcal{K}_{yyxy}=\mathcal{K}_{yyyx}=\mathcal{K}_{xxxy}=\mathcal{K}_{xxyx}=\mathcal{K}_{xyxx}=\mathcal{K}_{yxxx}=0$, and $\mathcal{K}_{xxyy}=\mathcal{K}_{xyxy}=\mathcal{K}_{xyyx}=\mathcal{K}_{yxxy}=\mathcal{K}_{yxyx}=\mathcal{K}_{yyxx}$. This means that there are three types of the intrinsic Kerr contributions for this mirror symmetric superconductor. In the above Lagrangian, we have kept the general $\zeta$ dependent form of the BCD Lagrangian to demonstrate that both forms result in the field dependent charge displacement in the Hamiltonian. Notice that such $\zeta$ dependence arises from a full-time derivative and, as such, can be gauged away, see detailed discussion below. In the main text we set $\zeta =0$.  

We now perform the standard procedure to find the circuit Hamiltonian, by defining the canonical charges (momentum) as:
\begin{equation}
    \hbar n_x=p_x=\frac{\partial \mathcal L}{\partial \dot\phi_x}=\frac{\hbar^2 \dot\phi_x}{4 E_C^x}+(1-\zeta)\hbar g_{\rm BCD}\phi_x\phi_y ,\qquad\hbar n_y=p_y=\frac{\partial \mathcal L}{\partial \dot\phi_y}=\frac{\hbar^2 \dot\phi_y}{4 E_C^y}-\frac{\zeta}{2}\hbar g_{\rm BCD}\phi_x^2.
\end{equation}

The corresponding circuit Hamiltonian can be found using Legendre transformation as $H = \sum_ip_i \dot{\phi}_i - \mathcal{L}$ as 
\begin{equation}\label{Hofzeta}
\begin{aligned}
    H =&~2E^x_C (n_x-(1-\zeta)g_{\rm BCD}\phi_x\phi_y)^2 + 2E^y_C (n_y+\frac{\zeta}{2}g_{\rm BCD}\phi_x^2)^2 \\&+\frac{E^x_{L}}{2} \phi_x^2+ \frac{E^y_{L}}{2} \phi_y^2 + \frac{\ell_x\ell_y}{4!}\left[\frac{\mathcal{K}_{xxxx}}{ L_x^4}\phi_x^4+\frac{6\mathcal{K}_{xxyy}}{ \ell_x^2 \ell_y^2}\phi_x^2\phi_y^2+\frac{\mathcal{K}_{yyyy}}{ \ell_y^4}\phi_y^4\right].
\end{aligned}
\end{equation}
In the above, $E_L^{x,y}$ is the total inductive energy originating from both the external circuit and the superconductor $E^x_{L}=E^x_{L,{\rm ext}}+E^x_{L,{\rm SC}},E^y_{L}=E^y_{L,{\rm ext}}+E^y_{L,{\rm SC}}$. Notice BCD element behaves as a nonlinear charge displacement that depends on $\phi_\alpha \phi_\beta$. We note that the $\zeta$ dependence in Eq.~(\ref{Hofzeta}) is a gauge and its dependence in the classical circuit Hamiltonian in Eq.~(\ref{Hofzeta}) can be eliminated by canonical transform. To see this, notice that the transformation below preserves the cannonical structure of the Hamiltonian: 
\begin{equation}
    \bar \phi_x = \phi_x,\quad \bar \phi_y= \phi_y, \quad \bar n_x=  n_x - \zeta g_{\rm BCD}\phi_x\phi_y,\quad \bar n_y= n_y - \frac{\zeta}{2} g_{\rm BCD}\phi^2_x,
\end{equation}
since $\{\phi_\alpha,\phi_\beta\}=0$ and $\{\bar n_x,\bar n_y\}= \{n_x,n_y\}-\zeta g_{\rm BCD}\phi_x \{\phi_y,n_y\}- \frac{\zeta}{2} g_{\rm BCD}\{n_x, \phi^2_x\}=\{n_x,n_y\}$. 

In quantizing the Hamiltonian we promote $\phi_x\rightarrow\hat{\phi}_m$, $\phi_y\rightarrow \hat{\phi}_b$ and $n_x\rightarrow \hat n_m$, $n_y\rightarrow \hat n_b$ to conjugate quantum operators satisfying $[\hat{\phi}_i,\hat{n}_j]=i\delta_{ij}$ and write 
\begin{equation}\label{Hquantzeta}
\begin{aligned}
    \hat H(\zeta) =&~2E^x_C (\hat n_m-(1-\zeta)g_{\rm BCD}\hat\phi_m\hat\phi_b)^2 + 2E^y_C (\hat n_b+\frac{\zeta}{2}g_{\rm BCD}\hat\phi_m^2)^2 \\&+\frac{E^x_{L}}{2} \hat\phi_m^2+ \frac{E^y_{L}}{2} \hat\phi_b^2 + \frac{\ell_x\ell_y}{4!}\left[\frac{\mathcal{K}_{xxxx}}{ \ell_x^4}\hat\phi_m^4+\frac{6\mathcal{K}_{xxyy}}{ \ell_x^2 \ell_y^2}\hat\phi_m^2\hat\phi_b^2+\frac{\mathcal{K}_{yyyy}}{ \ell_y^4}\hat\phi_b^4\right],
\end{aligned}
\end{equation}

Notice that this quantum circuit Hamiltonian is $\zeta$ dependent. Similar to the cannonical transformation for the classical case, here $\zeta$ dependence in the quantum circuit Hamiltonian in Eq. (\ref{Hquantzeta}) can be eliminated with a unitary transformation. To see this consider the unitary transformation: $\hat H(0) =\hat U^\dagger(\zeta) \hat H(\zeta)\hat U(\zeta)$, where $\hat U(\zeta) = \exp [-i\zeta g_{\rm BCD}\phi_m^2 \phi_b/2]$. Indeed $\hat U^\dagger(\zeta ) \hat\phi_\alpha \hat U(\zeta ) = \hat\phi_\alpha, \hat U^\dagger(\zeta ) \hat n_m \hat U(\zeta ) =n_m -g_{\rm BCD}\zeta \phi_b\phi_m , \hat U^\dagger \hat n_b \hat U =n_b-g_{\rm BCD}\zeta \phi_m^2/2$ where we used $[\hat \phi_\alpha,\hat n_\beta]=i\delta_{\alpha\beta}$ and the Baker-Campbell-Hausdorff formula so that 
\begin{equation}
    e^{X}Y e^{-X} = Y +[X,Y] +\frac{1}{2!}[X,[X,Y]]+\frac{1}{3!}[X,[X,[X,Y]]]+\ldots.
\end{equation}
This unitary transformation $\hat U(\zeta) = \exp [-i\zeta g_{\rm BCD}\phi_m^2 \phi_b/2]$ brings Eq.~(\ref{Hquantzeta}) to the same quantum circuit Hamiltonian but with $\zeta =0$. Hereafter we set $\zeta = 0$. By expanding in the linear order of $g_{\rm BCD}$ we $\hat H_{\rm BCD} =-2E^x_C g_{\rm BCD} (\hat n_m \hat\phi_m\hat\phi_b + \hat\phi_m\hat\phi_b \hat n_m)$. After substituting $\hat{\phi}_i=\varphi_{i,{\rm zpf}}(\hat a_i+\hat a_i^\dagger)$, $\hat n_i=i (\hat a_i^\dagger-\hat a_i)/(2\varphi_{i,{\rm zpf}})$ where $i\in(m,b)$, $\hat a_m = \hat m, \hat a_b =\hat b$, $\varphi_{b,{\rm zpf}}=(E^y_{C}/E^y_{L})^{1/4}$ and $\varphi_{m,{\rm zpf}}=(E^x_{C}/E^x_{L})^{1/4}$, we obtain Eq. (\ref{eq:sBCD_quantized_hamiltonian}) of the main text. 

Note that in three terminal set-up $E_C^x$ and $E_C^y$ are not linearly independent of each other, which can be seen from their definitions $E_{C}^y = e^2/[2C_0]$, $E_C^{x}=2e^2/(C_0+2C_d)$. This mean that for a given $E_C^y$, we have $E_C^x < 2E_C^y$. Moreover, we that for a circuit tuned to the autoparameteric condition $\omega_b = 2\omega_m$ we also have 
\begin{equation}
    E^{x}_L = \hbar^2\omega^2_m/(4E^x_C),\qquad E^{y}_L = \hbar^2\omega^2_m/E^x_C, \label{here0}
\end{equation}
Positiveness of $L_d$ further restricts the available values of $E_C^x$. To see this remember:
\begin{align}
    &E_L^x=(1/[2L_{0}]+1/L_d)(\hbar/[2e])^2 + E^x_{L,{\rm SC}}=\hbar^2\omega^2_m/(4E^x_C),\label{here1}\\ &E_L^y =(2/L_0)(\hbar/[2e])^2 + E^y_{L,{\rm SC}} =\hbar^2\omega^2_m/E^x_C,\label{here2}
\end{align}
where for the second equality on the RHS of Eq.~(\ref{here2}) we used Eq.~(\ref{here0}). By using Eq.~(\ref{here1}) we write $(1/L_d)(\hbar/[2e])^2= \hbar^2\omega_m^2/(4E_{C}^x)-E_{L,{\rm SC}}^x-(1/[2L_0])(\hbar/[2e])^2 \geq 0$. Then we use Eq.~(\ref{here2}) to eliminate ${L_0}$ term and find the allowed region as
\begin{equation}
    \hbar^2\omega_m^2/(E_C^x)-\hbar^2\omega_m^2/(E_C^y)\geq4E_{L,{\rm SC}}^x -E_{L,{\rm SC}}^y.
\end{equation}
The equality defines the boundary and was evaluated and displayed as dashed line on Fig.~\ref{sBCDQFIG3} of the main text for $E_{L,{\rm SC}}^x/h= E_{L,{\rm SC}}^y/h = 50$ GHz. From the inequality follows that the allowed region on the Fig.~\ref{sBCDQFIG3} for the 3T device is left with respect to the boundary. Interestingly, if $E_{L,{\rm SC}}^\alpha$ is neglected, the boundary becomes a straight line $E_{C}^x\leq E_{C}^y$.

\subsubsection{Four terminal device as shown in Fig.~\ref{fig:deviceSM}(b)} 

In this subsection, we consider a four-terminal geometry shown in Fig. \ref{fig:deviceSM}(b) with horizontal and vertical directions to be identified as $x$ and $y$ respectively. Notably such a device architecture is more transparent and the parameter space of possible auto-parametric resonances is larger yet as a trade-off such system generally hosts a quadrupole mode. Here we assume such is not exited. External loads in $x$ direction include $E^\alpha_{L}$ and $E^\alpha_{C}=e^2/C^\alpha$. We define: $\phi_{2,x}-\phi_{1,x} = \phi_x,\phi_{2,y}-\phi_{1,y} = \phi_y$ to find, for example, the circuit Lagrangian as:
\begin{align}
    {\mathcal{L}} \!=\! \frac{\hbar^2 \dot\phi_{x}^2}{8E^x_{C}} + \frac{\hbar^2 \dot\phi_{y}^2}{8E^y_{C}} - \frac{E^x_{L}}{2} \phi_{x}^2 - \frac{E^y_{L}}{2} \phi_{y}^2-U_{\rm SC}
\end{align}
where we used that $E^x_{L,SC}={A D^{xx}_s}/{\ell_x^2}$ and neglected terms of order $\mathcal{O}(\phi^5,\omega^2)$.  For the chosen circuit geometry, we write that the current that follows from the superconductor into LC branch as: $I_{x} = ({e\hbar }/{2E^x_{C}}) \ddot{\phi}_{x} + ({2e}/{\hbar}) E^x_{L} \phi_{x}.$ The corresponding full circuit Hamiltonian can be shown to be:
\begin{equation}
    {H} = 2 E^x_{C} \left(n_x - \frac{\mathcal{D}}{8\ell_x} \phi_x \phi_y\right)^2 + \frac{E^{x}_{L,\text{eff}}}{2} \phi_x^2 +2 E^y_{C} n_y^2 + \frac{E^{y}_{L,\text{eff}}}{2} \phi_y^2+ \frac{\ell_x\ell_y}{4!}\left[\frac{\mathcal{K}_{xxxx}}{ \ell_x^4}\phi_x^4+\frac{6\mathcal{K}_{xxyy}}{ \ell_x^2 \ell_y^2}\phi_x^2\phi_y^2+\frac{\mathcal{K}_{yyyy}}{ \ell_y^4}\phi_y^4\right].
\end{equation}

\section{Microscopic expressions for superfluid weight, BCD nonlinearity, and intrinsic Kerr nonlinearity}\label{QPandMat}

In this section, we provide microscopic expressions for the superfluid weight, BCD nonlinearity, and intrinsic Kerr nonlinearity. To do this we analyze the superconductor response by examining the superconducting state via a Bogoliubov-de-Gennes (BdG) mean field Hamiltonian $\hat{H}= \sum_{\mathbf{k}}\Psi^\dagger_{\mathbf{k,q}} \hat{H}^{\rm BdG}_{\mathbf{k,q}}[\mathbf{A}(t)]\Psi_{\mathbf{k,q}}$ in the Bloch-Nambu basis $\Psi_{\mathbf{k,q}}=(\vec{c}_{\mathbf{k}+\mathbf{q},\uparrow},\vec{c}^{~\dagger}_{-\mathbf{k}+\mathbf{q},\downarrow})^{\mathrm{T}}$:
\begin{equation}
    \hat{H}^{\rm BdG}_{\mathbf{k,q}}[\mathbf{A}_t] \hspace{-1mm}=\hspace{-1mm}{\left(\begin{array}{cc}
         \hat{H}_{\mathbf{k+q}}^{(0)}[\mathbf{A}_t]&\hat \Delta  \\
         \hat{\Delta}^\dagger&-\{\hat{H}^{(0)}_{\mathbf{-k+q}}[\mathbf{A}(t)]\}^T
    \end{array}\right)},
    \label{eq:BdGHSI}
\end{equation}
where $H_{\mathbf{k}}^{(0)} =  H_0(\mathbf{k}) - \mu\mathbb{I}$ is the the Bloch Hamiltonian and $\hat{\Delta}$ is the pairing potential. Here $\mu$ is the chemical potential and $H_0(\mathbf{k}) \ket{n, \mathbf{k}} = \epsilon_{\mathbf{k}} \ket{n, \mathbf{k}}$ with $\epsilon_{\mathbf{k}}$ and $\ket{n, \mathbf{k}}$ the energy and eigenstate respectiveily. Notice that coupling to electromagnetic $\mathbf{A}_t$ proceeds in the standard fashion through the kinetic part of the BdG Hamiltonian \cite{DeGennes}. As a result, $\partial_\mathbf{A} = (e/\hbar)\partial_\mathbf{q} = (e/\hbar)\tau_z\partial_\mathbf{k}$ where $\tau_z$ is the Pauli matrix acting in the Nambu basis \cite{sBCD_accepted, PhysRevB.96.064511}. 

To systematically derive the linear and nonlinear optical response of a superconductor in powers of $A$, we first note that the BdG Hamiltonian can be expanded as 
\begin{equation}\label{eq:hBdGexp}
    \hat{H}^{\rm BdG}_{\mathbf{k,q},t} = H^{\rm BdG}_{\mathbf{k},\mathbf{q}}+\frac{e}{\hbar}A^\alpha_t \partial_\mathbf{q}^\alpha \hat{H}^{\rm BdG}_{\mathbf{k},\mathbf{q}}+\frac{1}{2!}\frac{e^2}{\hbar^2}A^\alpha_t A^\beta_t \partial_\mathbf{q}^\alpha \partial_\mathbf{q}^\beta \hat{H}^{\rm BdG}_{\mathbf{k},\mathbf{q}}+\frac{1}{3!}\frac{e^3}{\hbar^3}A^\alpha_t A^\beta_t A^\gamma_t\partial_\mathbf{q}^\alpha \partial_\mathbf{q}^\beta \partial_\mathbf{q}^\gamma \hat{H}^{\rm BdG}_{\mathbf{k},\mathbf{q}}+\mathcal{O}(A^4_t).
\end{equation}
where $\hat{H}^{\rm BdG}_{\mathbf{k,q}}[0]\equiv \hat{H}^{\rm BdG}_{\mathbf{k,q}}$ with $\hat{H}^{\rm BdG}_{\mathbf{k,q}} \ket{\Psi_a(\mathbf{k,q})} = \mathcal{E}_{a,{\mathbf{k}}}(\mathbf{q})\ket{\Psi_a(\mathbf{k,q})}$ describing the BdG energies and BdG eigenstates. In what follows, we will omit explicit $\mathbf{k,q}$ dependence for brevity. The expansion coefficients can be written as $h^\alpha_{ab}=\bra{\Psi_a}(\partial_\mathbf{q}^\alpha H^{\rm BdG}_{\mathbf{k},\mathbf{q}})\ket{\Psi_b}$ and $h^{\alpha\gamma}_{ab}=\bra{\Psi_a}(\partial_\mathbf{q}^\alpha\partial_\mathbf{q}^\gamma H^{\rm BdG}_{\mathbf{k},\mathbf{q}})\ket{\Psi_b}$ in the BdG basis; these can be directly evaluated from the BdG energies $\mathcal{E}_{a,{\mathbf{k}}}$ and the BdG Berry connections $\mathcal{A}^\alpha_{ab} = \bra{\Psi_a}i\partial^\alpha_{\mathbf{q}}\Psi_b\rangle$ [see e.g., Ref.~\cite{sBCD_accepted,Yanase1, Yanase2}]. 

To obtain the responses, we will adopt an adiabatic turning on to regularize the electromagnetic fields $A^\alpha_t=e^{\eta t}\int_{\omega_1} A^\alpha_{\omega_1} e^{-i\omega_1 t}$ (here $\int_{\omega_1}\equiv \int_{-\infty}^\infty d\omega_1/(2\pi)$). We follow the standard perturbation theory and iteratively solve the quantum Liouville equation: $i\hbar \partial_t\hat{\rho}(t)= [\hat{H}^{\rm BdG}_{\mathbf{k,q}}[\mathbf{A}(t)],\hat{\rho}(t)],$ where $\hat\rho(t)$ is the time-dependent density matrix~\cite{sBCD_accepted,Yanase1, Yanase2}. using the BdG Hamiltonian expansion in Eq.(\ref{eq:hBdGexp}). This procedure is outlined in detail in Ref.~\cite{sBCD_accepted}. The current can be directly evaluated as $\hat{j}^\alpha(t)\equiv\partial \hat H ^{\rm BdG}_{\mathbf{k,q}}/ \partial A^\alpha_t$ with the density matrix: $j^\gamma(t) = \int_{\mathbf{k}}{\rm Tr}[\hat{j}^\gamma(t) \hat{\rho}(t)]$ where $\int_{\mathbf{k}}\equiv\int{d\mathbf{k}}/{(2\pi)^2}$. 

At linear order in perturbation theory, we have $j_{(1)}^\alpha(t) =    \frac{e}{\hbar} \int_{\omega_1} \rho^{\alpha\beta}_{s,\omega_1} q^\beta_{\omega_1}$ where
\begin{equation}\label{SIconductivity}
    \rho^{\alpha\beta}_{s,\omega_1}=\int_{\mathbf{k}}\left[\sum_a f_a\partial^\beta h^\alpha_{aa} + \sum_{ab}\frac{(\hbar\omega_1+i\eta)h^\beta_{ab}h^\alpha_{ba}f_{ab}}{\hbar\omega_1 -\mathcal{E}_{ab}+ i\eta}\frac{1}{\mathcal{E}_{ba}}\right],
\end{equation}
where $\mathcal{E}_{ba}= \mathcal{E}_{\mathbf{k},a}(\mathbf{q})-\mathcal{E}_{\mathbf{k},a}(\mathbf{q}), f_{ab}= f[\mathcal{E}_{\mathbf{k},a}(\mathbf{q})]-f[\mathcal{E}_{\mathbf{k},b}(\mathbf{q})]$ and $f [x]= 1/(1+\exp[x/(k_B T)])$. In obtaining Eq.~(\ref{SIconductivity}) we have noted $\mathbf{A}_t = (\hbar/e) \mathbf{q}_t$. The above is consistent with Ref.~\cite{PhysRevLett.123.237002}. Notice (as a sanity check) in the limit $\omega_1\rightarrow0$ that Eq.~(\ref{SIconductivity}) reduces to the familiar $\rho^{\alpha\beta}_{s,\omega_1}\equiv\rho^{\alpha\beta}_{s}=\int_{\mathbf{k}}\sum_a f_a\partial^\beta_\mathbf{q}\partial^\alpha_\mathbf{q} \mathcal{E}_{a,\mathbf{k}}(\mathbf{q})$. We will employ this formula hereafter. Now, we construct the superfluid weight Lagrangian from the current density using  Eq.~(\ref{s3}) [note, $I_x = \ell_y j_x= (\ell_x\ell_y/\ell_x)j_x$ or $I_\alpha = (\ell_x\ell_y/\ell_\alpha)j^\alpha$] to find:
\begin{equation}
     \mathcal{L}_{\rm sw} = -\frac{1}{2}\frac{\ell_x\ell_y}{ \ell_\alpha \ell_\beta} (\frac{1}{4} \rho^{\alpha\beta}_{s})\phi_\alpha(t) \phi_\beta(t) 
\end{equation}
with superfluid tensor as $D^{\alpha\beta}_{s}= \rho^{\alpha\beta}_{s}/4$.  Immediately we can write general inductive energy of the superconductor as $U_{\rm L, SC}=E^{\alpha\beta}_{\rm L,SC}\phi_\alpha \phi_\beta /2$ with $E^{\alpha\beta}_{\rm L,SC} = \ell_x\ell_y/({[\ell_\alpha \ell_\beta]})D^{\alpha\beta}_{s}$. In the main text we assumed that superfluid tensor is diagonal and square sample rendering $E^{\alpha\alpha}_{\rm L,SC} \equiv E^{\alpha}_{\rm L,SC}= D^{\alpha\alpha}_{s}$. 

As shown in Ref.\cite{sBCD_accepted}, the projected reduced BdG model can be written, after employing time-reversal symmetry, as the reduced BdG equation
\begin{equation}\label{reducedBdG}
    \left(\begin{array}{cc}
        \xi_{\mathbf{k+q}} & \Delta(\mathbf{k,q}) \\
        \Delta^*(\mathbf{k,q}) & -\xi_{\mathbf{k-q}}
    \end{array}\right)\left(\begin{array}{c}
         \bar{\mathcal{U}}_{\mathbf{k},\mathbf{q}}(\mathbf{A})  \\
         \bar{\mathcal{V}}_{\mathbf{k},\mathbf{q}}(\mathbf{A})
    \end{array}\right)=\mathcal{E}_{\mathbf{k}}(\mathbf{q})\left(\begin{array}{c}
         \bar{\mathcal{U}}_{\mathbf{k},\mathbf{q}}(\mathbf{A})  \\
         \bar{\mathcal{V}}_{\mathbf{k},\mathbf{q}}(\mathbf{A})
    \end{array}\right),
\end{equation}
where $\Delta_{\mathbf{k,q}} = \bra{n_{\rm FS},\mathbf{k+q}}\hat{\Delta}\ket{n_{\rm FS},\mathbf{k-q}}$ and $\xi_\mathbf{k} = \epsilon_{\mathbf{k}}-\mu$, assuming that the Fermi energy $\mu$ crosses the conduction band with dispersion $\epsilon_{\mathbf{k}}$. We explicitly keep finite $\mathbf{q}$ in order to track the coupling to the vector potential; the final limit of interest is $\mathbf{q}\rightarrow0$. Importantly, the $\mathbf{k,q}$ dependence of the superconducting gap $\Delta_{\mathbf{k},\mathbf{q}}(\mathbf{A})$ arises from the projection of the multiband Hamiltonian onto shifted Bloch states at the Fermi surface, $\ket{n_{\rm FS},\pm\mathbf{k+q}}$. 

\vspace{2mm}

\paragraph{Superfluid stiffness.} Here we approximate the superfluid stiffness in the isolated-band approximation for a time-reversal-symmetric superconductor. To do this we remind that the lower Bogoliubov-band energy is $2\mathcal{E}_{\mathbf{k},-}(\mathbf{q}) = \epsilon_{\mathbf{k+q}}-\epsilon_{\mathbf{k-q}}- [(\tilde{\xi}_{\mathbf{k},\mathbf{q}})^2+4|\Delta_{\mathbf{k},\mathbf{q}}|^2]^{1/2}$ where $\tilde{\xi}_{\mathbf{k},\mathbf{q}} = \xi_{\mathbf{k+q}}+\xi_{\mathbf{k-q}}$ and $\xi_{\mathbf{k}}=\epsilon_\mathbf{k}-\mu$. Next we expand $\tilde{\xi}_{\mathbf{k},\mathbf{q}} = 2\xi_{\mathbf{k,0}}+q^\alpha q^\beta\partial_\mathbf{k}^\alpha\partial_\mathbf{k}^\beta\epsilon_{\mathbf{k}}+\mathcal{O}(\mathbf{q}^4)$ and $ |\Delta_{\mathbf{k,q}}|^2 =  |\Delta_{\mathbf{k,0}}|^2 -4 q^\alpha q^\beta \mathcal{M}_{\mathbf{k}}^{\alpha\beta}+\mathcal{O}(\mathbf{q}^4)$, where $\mathcal{M}_{\mathbf{k}}^{\alpha\beta}$ is the expansion coefficient, we obtain at $T\rightarrow0$ the superfluid weight as 
\begin{equation}\label{generic_stiffness}
    \begin{aligned}
    D^{\alpha \gamma}_s&= \frac{1}{4}\lim_{\mathbf{q}\rightarrow0}\int_{\mathbf{k}}\partial_\mathbf{q}^\gamma\partial^\alpha_{\mathbf{q}} \mathcal{E}_{\mathbf{k},-}(\mathbf{q})\\&= -\frac{1}{4}\lim_{\mathbf{q}\rightarrow0} \int_{\mathbf{k}}\partial_\mathbf{q}^\gamma\partial^\alpha_{\mathbf{q}} \sqrt{\xi_{\mathbf{k},0}^2+q^\alpha q^\beta \xi_{\mathbf{k},0}\partial_\mathbf{k}^\alpha\partial_\mathbf{k}^\beta\xi_{\mathbf{k},0}+ |\Delta_{\mathbf{k,0}}|^2 -4q^\alpha q^\beta \mathcal{M}^{\alpha\beta}_{\mathbf{k}}}\\&=-\frac{1}{4}\int_{\mathbf{k}} \frac{\xi_{\mathbf{k},0}\partial_\mathbf{k}^\alpha\partial_\mathbf{k}^\beta\xi_{\mathbf{k},0}}{\sqrt{\xi_{\mathbf{k},0}^2 + |\Delta_{\mathbf{k,0}}|^2}} +\int_{\mathbf{k}} \frac{\mathcal{M}^{\alpha\beta}_\mathbf{k}}{\sqrt{\xi_{\mathbf{k},0}^2 + |\Delta_{\mathbf{k,0}}|^2}}.
    \end{aligned}
\end{equation}
The details of the pairing are crucial for determining the proper superfluid stiffness of the material. For simplicity, in what follows we choose structureless pairing $\hat\Delta \sim \Delta_0\mathbb{I}$ that can be achieved in a simple proximitized platform with $\mathcal{M}_{\mathbf{k}}^{\alpha\beta}=|\Delta_0|^2g_{\mathbf{k}}^{\alpha\beta} =|\Delta_0|^2 {\rm Re}[\langle \partial^\alpha_\mathbf{k}n_{\rm FS},\mathbf{k}|(1-|n_{\rm FS},\mathbf{k}\rangle\langle n_{\rm FS},\mathbf{k}|)|\partial^\beta_\mathbf{k} n_{\rm FS},\mathbf{k}\rangle]$, where $g_{\mathbf{k}}^{\alpha\beta}$ is the quantum metric. 

\paragraph{Berry curvature dipole.} Now we estimate the BCD of time reversal symmetric superconductor within the isolated band approximation for structureless pairing potential. Such BCD is given only by the parent-band contribution, that in the limit of small gap reduces as follows:
\begin{equation}\label{SBCDIBA}
\mathcal{D}^{\alpha\beta\gamma}_{\rm BCD}=  2\int_{\mathbf{k}}|v_{\mathbf{k,0}}|^2 (\partial^\alpha_{\mathbf{k}} \partial^\beta_{\mathbf{k}} r_{n}^\gamma (\mathbf{k})-\partial^\alpha_{\mathbf{k}} \partial^\gamma_{\mathbf{k}} r_{n}^\beta (\mathbf{k}))=\int_{\mathbf{k}} \frac{ ( \partial^\beta_{\mathbf{k}} r_{n}^\gamma (\mathbf{k})-\partial^\gamma_{\mathbf{k}} r_{n}^\beta (\mathbf{k})) v^\alpha_\mathbf{k}|\Delta_0|^2}{((\epsilon_{\mathbf{k}}-\mu)^2+|\Delta_0|^2)^{3/2}},
\end{equation}
where $ r_{n}^\beta = \bra{n_{\rm FS},\mathbf{k}} i\partial_\mathbf{k}^\beta \ket{n_{\rm FS},\mathbf{k}}$ is the Berry connection of the conduction band.  Factor of 2 counts for spin. In obtaining the final form of Eq.~(\ref{SBCDIBA}), we used integration by parts. Equations~(\ref{generic_stiffness}) and~(\ref{SBCDIBA}) provide a quick material-estimation pipeline for assessing suitability for practical applications. 

\paragraph{Intrinsic Kerr effect.} In this subsection, we explicitly derive the intrinsic Kerr effect of a time-reversal-symmetric superconductor. We begin by considering the current density induced by an adiabatically varying condensate center of mass momentum $\mathbf{q}(t)$. To the leading order in $\mathbf{q}$ and frequency of the drive, the current is:
\begin{equation}
    j^\gamma(t) = \frac{e}{\hbar}\rho^{\gamma \alpha} q^\alpha + \tilde{\chi}^{\gamma\alpha\beta} q^\alpha \dot q^\beta+ \frac{e}{6\hbar}\bar{\mathcal{K}}^{\gamma\alpha\beta\delta} q^\alpha q^\beta q^\delta+\mathcal{O}(q^3\omega,q^5,\omega^2),
\end{equation}
where $\bar{\mathcal{K}}^{\gamma\alpha\beta\delta} =(1/V) \sum_{\mathbf{k}} f_a\partial^\gamma_\mathbf{q}\partial^\alpha_\mathbf{q}\partial^\beta_\mathbf{q}\partial_\mathbf{q}^\delta\mathcal{E}_{a,\mathbf{k}}(\mathbf{q})$. To find the strength of each term at the level of the energy or Hamiltonian we similarly perform the procedure applied to the superfluid weight, where we use Eq.~(\ref{s3}) to restore the corresponding Lagrangian of the superconductor as:
\begin{equation}
\begin{aligned}
    \mathcal{L}_{\rm SC} &=-\frac{1}{2}\ell_x\ell_y\frac{\hbar}{e}\frac{e}{\hbar}\rho^{\gamma \alpha}q^\gamma q^\alpha - \frac{1}{3}\ell_x\ell_y\frac{\hbar}{e}\tilde{\chi}^{\gamma\alpha\beta} q^\gamma q^\alpha \dot q^\beta- \frac{1}{4!}\ell_x\ell_y\frac{\hbar}{e}\frac{e}{\hbar}\bar{\mathcal{K}}^{\gamma\alpha\beta\delta} q^\gamma q^\alpha q^\beta q^\delta\\&=-\frac{1}{2}\frac{\ell_x\ell_y}{\ell_\gamma \ell_\alpha}\frac{1}{4}\rho^{\gamma \alpha}\phi_\gamma \phi_\alpha - \frac{\ell_x\ell_y}{\ell_\gamma \ell_\alpha \ell_\beta}\frac{e\tilde{\chi}^{\gamma\alpha\beta}}{24\hbar } \phi_\gamma\phi_\alpha \dot \phi_\beta- \frac{1}{4!}\frac{\ell_x\ell_y}{\ell_\gamma \ell_\alpha \ell_\beta \ell_\delta }\frac{\bar{\mathcal{K}}^{\gamma\alpha\beta\delta}}{16} \phi_\alpha \phi_\beta \phi_\gamma \phi_\delta.
\end{aligned}
\end{equation}
In is convenient to define ${\mathcal{K}}^{\gamma\alpha\beta\delta}=\bar{\mathcal{K}}^{\gamma\alpha\beta\delta}/16$ used in the main text. 

Verification of the Kerr and superfluid weight terms to reproduce the corresponding currents is conducted by direct substitution into the Euler-Lagrange equation. Since both depend only on $\mathbf{q}$ and contracted with symmetric tensors, one immediately reproduces the desired result. The consistency of the $\tilde \chi$ term in the with the current density and our previous BCD discussion, however, requires a bit more attention. Here we perform explicit substitution into the Euler-Lagrange equation and use Eq.~(\ref{s3}): 
\begin{equation}
     \mathcal{L}_{\tilde\chi}=- \frac{\ell_x\ell_y}{3}\frac{\hbar}{e}\tilde{\chi}^{\gamma\alpha\beta} q^\gamma q^\alpha \dot q^\beta,\quad\rightarrow\quad  \frac{\ell_x\ell_y\hbar}{e} j^\beta=\frac{d}{dt}\frac{\partial \mathcal{L}_{\tilde \chi}}{\partial \dot q^\beta} -\frac{\partial  \mathcal{L}_{\tilde \chi}}{\partial q^\beta} = \frac{\ell_x\ell_y}{3}\frac{\hbar}{e}(\tilde{\chi}^{\beta\alpha\gamma}+ \tilde{\chi}^{\alpha\beta\gamma}-\tilde{\chi}^{\gamma\alpha\beta}-\tilde{\chi}^{\alpha\gamma\beta})q^\alpha  \dot q^\gamma.
\end{equation}
Now, we employ that $\tilde \chi$ is gyrotropic $\tilde{\chi}^{\beta\alpha\gamma}=-\tilde{\chi}^{\gamma\alpha\beta}$, which in 2D also implies $\tilde \chi^{\alpha\beta\gamma} + \tilde\chi^{\beta\gamma\alpha} = \tilde\chi^{\beta\alpha\gamma}$ (since $(\alpha,\beta,\gamma)\in\{x,y\}$), thus we find $d_t({\partial \mathcal{L}_{\tilde \chi}}/{\partial \dot q^\beta}) -({\partial  \mathcal{L}_{\tilde \chi}}/{\partial q^\beta}) = -\ell_x\ell_y(\hbar/e)\tilde{\chi}^{\beta\alpha\gamma}q^\alpha  \dot q^\gamma$. Finally, assuming mirror $\mathcal{M}_x$ we have only two non-zero elements $\tilde \chi^{xxy} =e\mathcal{D}$ and $\tilde \chi^{yxx} =-e\mathcal{D}$, e.g. $\mathcal{L}_{\tilde \chi}=- ({\hbar \ell_x\ell_y \mathcal{D} }/{3})((q^x)^2 \dot q^y -q^x \dot q^x q^y)=-({\hbar \ell_x\ell_y \mathcal{D} }/{3})(d[(q^x)^2 q^y]/dt -3q^x \dot q^x q^y)$, where full time derivative is physically irrelevant.

We now proceed to evaluate the Kerr tensor. We explicitly write the expansions needed to evaluate $\mathcal{E}_{\mathbf{k},-}(\mathbf{q})$ up to the fourth order in $\mathbf{q}$. Remember $2\mathcal{E}_{\mathbf{k},-}(\mathbf{q}) = \epsilon_{\mathbf{k+q}}-\epsilon_{\mathbf{k-q}}- [(\tilde{\xi}_{\mathbf{k},\mathbf{q}})^2+4|\Delta_{\mathbf{k},\mathbf{q}}|^2]^{1/2}$ where $\tilde{\xi}_{\mathbf{k},\mathbf{q}} = \xi_{\mathbf{k+q}}+\xi_{\mathbf{k-q}}$, $\xi_\mathbf{k} = \epsilon_{\mathbf{k}}-\mu$ and $|\Delta_{\mathbf{k,q}}|^2 = |\Delta_0|^2|\bra{n_{\rm FS},\mathbf{k+q}}{n_{\rm FS},\mathbf{k-q}}\rangle|^2$, thus:
\begin{align}
    & \tilde{\xi}_{\mathbf{k},\mathbf{q}} = 2\xi_{\mathbf{k,0}}+q^\alpha q^\beta\partial_\mathbf{k}^\alpha\partial_\mathbf{k}^\beta\epsilon_{\mathbf{k}}+\frac{1}{12}q^\alpha q^\beta q^\gamma q^\delta \partial_\mathbf{k}^\alpha\partial_\mathbf{k}^\beta \partial_\mathbf{k}^\gamma \partial_\mathbf{k}^\delta\epsilon_{\mathbf{k}} +\mathcal{O}(\mathbf{q}^6),\\
    & F(q) = 1 +\frac{q^\alpha q^\beta}{2} \partial_\mathbf{q}^\alpha\partial_\mathbf{q}^\beta F(\mathbf{q})\bigg|_{\mathbf{q}=0} +\frac{q^\alpha q^\beta q^\gamma q^\delta}{4!} \partial_\mathbf{q}^\alpha\partial_\mathbf{q}^\beta \partial_\mathbf{q}^\gamma\partial_\mathbf{q}^\delta F(\mathbf{q})\bigg|_{\mathbf{q}=0} + \mathcal{O} (q^6),
\end{align}
where $F(\mathbf{q}) = |\bra{n_{\rm FS},\mathbf{k+q}} n_{\rm FS},\mathbf{k-q}\rangle|^2 = {\rm Tr}[P_{\mathbf{k+q}}P_{\mathbf{k-q}}]$ for $P_\mathbf{k}\equiv \ket{n_{\rm FS},\mathbf{k}}\bra{n_{\rm FS},\mathbf{k}}$. Next, one can show:
\begin{equation}
    F(\mathbf{q}) =1+{\rm Tr} \bigg[q^\alpha q^\beta (P _\mathbf{k}P^{\alpha\beta}_\mathbf{k}-P^\alpha_\mathbf{k} P^\beta_\mathbf{k}) + q^\alpha q^\beta q^\gamma q^\delta \left(\frac{2}{4!}P_\mathbf{k} P_{\mathbf{k}}^{\alpha\beta\gamma\delta} -\frac{2}{3!} P^\delta_\mathbf{k}P^{\alpha\beta\gamma}_\mathbf{k} + \frac{1}{(2!)^2}P^{\alpha\beta}_\mathbf{k}P^{\gamma\delta}_\mathbf{k}\right)\bigg] + \mathcal{O}(q^6),
\end{equation}
where $P^{\alpha\beta\ldots}_\mathbf{k} \equiv (\partial_\mathbf{k}^\alpha  \partial_\mathbf{k}^\beta \ldots) P_\mathbf{k}$. Above can be simplified due to ${\rm Tr}[P_\mathbf{k}^2] = 1 ,{\rm Tr}[P_\mathbf{k}P_\mathbf{k}^\alpha]=0,{\rm Tr}[P_\mathbf{k}P_\mathbf{k}^{\alpha\beta}]=-{\rm Tr}[P_\mathbf{k}^{\alpha}P^\beta_\mathbf{k}], q^\alpha q^\beta q^\gamma q^\delta{\rm Tr}[P^{\alpha\beta\gamma\delta}_\mathbf{k}P_\mathbf{k} +4P^{\alpha\beta\gamma}_\mathbf{k}P^\delta_\mathbf{k}+ 3 P^{\alpha\beta}_\mathbf{k}P^{\delta\gamma}_\mathbf{k}] =0$, to obtain
\begin{equation}
    F(\mathbf{q}) = 1-2q^\alpha q^\beta{\rm Tr} [P^\alpha_\mathbf{k} P^\beta_\mathbf{k}] - \frac{2}{3}q^\alpha q^\beta q^\gamma q^\delta {\rm Tr}[ P^\delta_\mathbf{k}P^{\alpha\beta\gamma}_\mathbf{k} ].
\end{equation}
Next we find:
\begin{multline}
[(\tilde{\xi}_{\mathbf{k},\mathbf{q}}/2)^2+|\Delta_{\mathbf{k},\mathbf{q}}|^2]^{1/2} =\sqrt{\xi_{\mathbf{k,0}}^2+\Delta_0^2} [1+q^\alpha q^\beta A^{\alpha\beta}_\mathbf{k}+q^\alpha q^\beta q^\gamma q^\delta B^{\alpha\beta\gamma\delta}_\mathbf{k}  ]^{1/2}+ \mathcal{O}(q^6)=\\=\sqrt{\xi_{\mathbf{k,0}}^2+\Delta_0^2} \left[1+\frac{1}{2}q^\alpha q^\beta A^{\alpha\beta}_\mathbf{k}+q^\alpha q^\beta q^\gamma q^\delta \left(\frac{1}{2}B^{\alpha\beta\gamma\delta}_\mathbf{k}  -\frac{1}{8}A^{\alpha\beta}_\mathbf{k}A^{\gamma\delta}_\mathbf{k}\right)\right]+ \mathcal{O}(q^6).
\end{multline}
where
\begin{align}
    &A^{\alpha\beta}_\mathbf{k} = \frac{\xi_{\mathbf{k,0}}\partial_\mathbf{k}^\alpha\partial_\mathbf{k}^\beta\xi_{\mathbf{k,0}}-2\Delta_0^2{\rm Tr} [P^\alpha_\mathbf{k} P^\beta_\mathbf{k}]}{\xi_{\mathbf{k,0}}^2+\Delta_0^2}\\
    &B^{\alpha\beta\gamma\delta}_\mathbf{k} =\frac{1}{12}\frac{(\xi_{\mathbf{k,0}}\partial_\mathbf{k}^\alpha\partial_\mathbf{k}^\beta \partial_\mathbf{k}^\gamma \partial_\mathbf{k}^\delta\xi_{\mathbf{k,0}}+3(\partial_\mathbf{k}^\alpha\partial_\mathbf{k}^\beta\xi_{\mathbf{k,0}})(\partial_\mathbf{k}^\gamma\partial_\mathbf{k}^\delta\xi_{\mathbf{k,0}})}{\xi_{\mathbf{k,0}}^2+\Delta_0^2} - \frac{2}{3}\frac{\Delta_0^2}{\xi_{\mathbf{k,0}}^2+\Delta_0^2} {\rm Tr}[ P^\delta_\mathbf{k}P^{\alpha\beta\gamma}_\mathbf{k} ]).
\end{align}

Thus using ${\mathcal{K}}^{\gamma\alpha\beta\delta} =(1/[16V]) \lim_{\mathbf{q}\rightarrow0}\sum_{\mathbf{k}} \partial^\gamma_\mathbf{q}\partial^\alpha_\mathbf{q}\partial^\beta_\mathbf{q}\partial_\mathbf{q}^\delta\mathcal{E}_{\mathbf{k},-}(\mathbf{q})$, we find
\begin{equation}\label{Kfull}
    {\mathcal{K}}^{\gamma\alpha\beta\delta} =-\frac{3}{4}\int_\mathbf{k} \sqrt{\xi_{\mathbf{k,0}}^2+\Delta_0^2} \left(B^{\alpha\beta\gamma\delta}_\mathbf{k}  -\frac{1}{4}A^{\alpha\beta}_\mathbf{k}A^{\gamma\delta}_\mathbf{k}\right) = {\mathcal{K}}^{\gamma\alpha\beta\delta}_1+ {\mathcal{K}}^{\gamma\alpha\beta\delta}_2+ {\mathcal{K}}^{\gamma\alpha\beta\delta}_3+ {\mathcal{K}}^{\gamma\alpha\beta\delta}_4 
\end{equation}
where
\begin{align}
&{\mathcal{K}}^{\gamma\alpha\beta\delta}_{1} =-\frac{1}{4}\int_\mathbf{k} \frac{\Delta_0^2 \partial^\alpha_\mathbf{k}\partial^\beta_\mathbf{k}\xi_{\mathbf{k,0}}}{(\xi_{\mathbf{k,0}}^2+\Delta_0^2)^{3/2} } \partial_\mathbf{k}^\gamma \partial_\mathbf{k}^\delta\epsilon_{\mathbf{k}}+\frac{3}{16}\int_\mathbf{k} \xi_{\mathbf{k,0}}\frac{\Delta_0^2 (\partial^\alpha_\mathbf{k}\xi_{\mathbf{k,0}})(\partial^\beta_\mathbf{k}\xi_{\mathbf{k,0}})}{(\xi_{\mathbf{k,0}}^2+\Delta_0^2)^{5/2} } \partial_\mathbf{k}^\gamma \partial_\mathbf{k}^\delta\epsilon_{\mathbf{k}},\\
    &{\mathcal{K}}^{\gamma\alpha\beta\delta}_2 =-\frac{3}{4}\int_\mathbf{k}  \frac{\xi_{\mathbf{k,0}}\Delta_0^2}{(\xi_{\mathbf{k,0}}^2+\Delta_0^2)^{3/2}}{\rm Tr} [P^\gamma_\mathbf{k} P^\delta_\mathbf{k}]\partial_\mathbf{k}^\alpha\partial_\mathbf{k}^\beta\epsilon_{\mathbf{k}}
    \\
    &{\mathcal{K}}^{\gamma\alpha\beta\delta}_3 =-\frac{3}{4}\int_\mathbf{k} \left(\frac{2}{3}\frac{\Delta_0^2}{(\xi_{\mathbf{k,0}}^2+\Delta_0^2)^{1/2}} {\rm Tr}[  P^{\alpha\delta}_\mathbf{k}P^{\beta\gamma}_\mathbf{k}]) -\frac{\Delta_0^4}{(\xi_{\mathbf{k,0}}^2+\Delta_0^2)^{3/2}} {\rm Tr} [P^\alpha_\mathbf{k} P^\beta_\mathbf{k}]{\rm Tr} [P^\gamma_\mathbf{k} P^\delta_\mathbf{k}]\right)
    \\
    &{\mathcal{K}}^{\gamma\alpha\beta\delta}_{4} =\frac{1}{2}\int_\mathbf{k} \frac{\Delta_0^2 \xi_{\mathbf{k,0}}\partial^\alpha_\mathbf{k}\xi_{\mathbf{k,0}}}{(\xi_{\mathbf{k,0}}^2+\Delta_0^2)^{3/2}} {\rm Tr}[ P^\delta_\mathbf{k}P^{\beta\gamma}_\mathbf{k}].
\end{align}
Notice, for the system with mirror $\mathcal{M}_x$ axis, the fully symmetric tensor $\hat{\mathcal{K}}$ has only three independent components $\mathcal{K}^{xxxx}, \mathcal{K}^{yyyy},\mathcal{K}^{xxyy}=\mathcal{K}^{yyxx}=\mathcal{K}^{yxyx}=\mathcal{K}^{xyxy}=\mathcal{K}^{xyyx}=\mathcal{K}^{yxxy}$.

\paragraph{Scaling of circuit quantum nonlinearities}. 
As discussed in the main text, the nonlienarities can be quantized. To see this, we write $E^\alpha_{C} = \hbar \omega_\alpha \phi^2_{\alpha,{\rm zpf}}/2$, $\hbar\omega_\alpha = \sqrt{4 E^\alpha_{C}E^{\alpha}_{L,\text{eff}}}$ and $\phi_{\alpha,{\rm zpf}} = (E^\alpha_{C}/E^{\alpha}_{L,\text{eff}})^{1/4}$ and find the quantized circuit Hamiltonian as:
\begin{equation}
    \hat H \!=\! \hbar\omega_m (\hat m^\dagger \hat m +\frac{1}{2}) + \hbar\omega_b (\hat b^\dagger \hat b +\frac{1}{2}) - i \hbar g_{\rm BCD} (\hat m^{\dagger 2}\!-\hat m^2)(\hat b+\hat b^\dagger) +U_m(\hat m+\hat m^\dagger)^4\! +U_b(\hat b+\hat b^\dagger)^4\!+U_{mb}(\hat b+\hat b^\dagger)^2 (\hat m+\hat m^\dagger)^2, 
\end{equation}
where $U_m = A\mathcal{K}_{xxxx}\phi_{x,{\rm zpf}}^4/(4! \ell_x^4),U_b = A\mathcal{K}_{yyyy}\phi_{y,{\rm zpf}}^4/(4! \ell_y^4)$ and the cross Kerr effect has two contributions $U_{mb} = U_{mb}^{I}+U_{mb}^{D}$ where $U^I_{mb} = A\mathcal{K}_{xxyy}\phi_{x,{\rm zpf}}^2\phi_{y,{\rm zpf}}^2/(4 \ell_x^2 \ell_y^2)$ and $U^D_{mb} = E_C^x \mathcal{D}^2\phi_{x,{\rm zpf}}^2\phi_{y,{\rm zpf}}^2/(32 \ell_x^2)$. This description applicable to both three and four terminal devices. However, for three terminal devices, the permissive values of $E_C$ and $E_L$ must be employed. This is also discussed in Eqs. (\ref{here0}-\ref{here2}) of section B. 

It is useful to understand the scaling of the strength of the nonlinearities at quantum circuit level. To do so we focus on auto-parametric resonance regime $\omega_b = 2\omega_m$ considered in the main text. Notice after normal ordering within the rotating wave approximation, we obtain $(\hat a^\dagger_i +\hat a)^4\rightarrow6 \hat a^{\dagger2}_i\hat  a^2_i + 12\hat a^\dagger \hat a +  3,(\hat a^\dagger_i +\hat a)^2\rightarrow2\hat a^\dagger_i \hat a_i +  1$. The number terms renormalize the mode frequencies as  $\tilde{\omega}_m = \omega_m + (12U_m+2U_{mb})/\hbar$ and $\tilde{\omega_b} = \omega_b + (12U_b+2U_{mb})/\hbar$. Thus the auto-parametric resonance is defined as $\tilde\omega_b = 2\tilde\omega_m$ (in the main text, for simplicity, we ignored renormalization of the frequencies). Within the rotating wave approximation, the quantum three-wave and four-wave mixing terms read as: 
\begin{equation}
    \hat H' \!=\!  - i \hbar g_{\rm BCD} (\hat m^{\dagger 2}\hat b\!-\hat m^2 \hat b^\dagger)+\hbar K_x \hat m^{\dagger 2}\hat m^2 + \hbar (K_{xy}+K^D_{xy})\hat m^\dagger  \hat m\hat b^\dagger\hat b + \hbar K_y \hat b^{\dagger 2}\hat b^2, 
\end{equation}
with coefficients $g_{\rm BCD}$ as in the main text and
\begin{align}
    &\hbar K_{x}\! = \!\frac{\ell_y \mathcal{K}_{xxxx}}{4  \ell_x^3} \phi_{{\rm zpf},m}^4=\frac{\ell_y \mathcal{K}_{xxxx}}{ \ell_x^3} \frac{(E_C^x)^2}{\hbar^2\omega_m^2},\quad \hbar K_{y}\! = \!\frac{\ell_x \mathcal{K}_{yyyy}}{4  \ell_y^3} \phi_{{\rm zpf},b}^4=\frac{\ell_x \mathcal{K}_{yyyy}}{\ell_y^3} \frac{(E_C^y)^2}{\hbar^2\omega_b^2}\\
    & \hbar K^{D}_{xy} \!=\!   \frac{E^x_{C}\mathcal{D}^2}{8 \ell_x^2}\phi^2_{{\rm zpf}, b}\phi^2_{{\rm zpf}, m},\qquad  \hbar K_{xy}\!=\!  \frac{ \mathcal{K}_{xxyy}}{ \ell_x \ell_y}\phi_{{\rm zpf},m}^2 \phi_{{\rm zpf},b}^2=\frac{4 \mathcal{K}_{xxyy}}{ \ell_x\ell_y} \frac{E_C^x E_C^y}{\hbar\omega_m \hbar\omega_b}
\end{align}
where $K^{D}_{xy}$ is small due to both smallness of $\mathcal{D}/\ell_x$ and zpf. The above set of equations is consistent with Eq.~(\ref{APKERR}) of the main text. Notice that device geometry and specific inductor and capacitor loads allow multiple parameters to control the relative strength of BCD and Kerr effects. For example larger size suppresses Kerr effects as $\ell^{-2}$ while BCD mixing goes as $\ell^{-1}$. Also, as shown on Fig.~(\ref{sBCDQFIG3}) of the main text, for larger $\omega_m$ the BCD mixing (scales as $\sim \omega_m^{-1/2}$), and can become stronger then the Kerr nonlinearity (scales as $\omega_m^{-2}$) as discussed in the main text. 

\section{Numerical estimates of nonlinearities in a superconducting proximitised twisted bilayer graphene on hexagonal Boron Nitride platform}

In this section, we discuss our numerical simulations. We acknowledge the use of ChatGPT (OpenAI) GPT-5.6 Sol and GPT 6.1 Sol in debugging, optimisation, and stress testing. All simulations were verified and checked line-by-line by author. 

Below we discuss details of our numerical estimates of the BCD mixing in twisted bilayer graphene on hexagonal Boron Nitride structures. For the superconducting circuit, we will focus on proximtized TBG/hBN. For simplicity, we will use a structureless pairing potential when computing the BdG states used for estimating the BCD and Kerr effects, see discussion in Ref.~\cite{sBCD_accepted}. 

To simulate the TBG/hBN parent bandstructure, we employ the Bistritzer-MacDonald model \cite{BMmodel,PhysRevX.8.031087}. To capture the BCD effect which requires broken inversion and rotational symmetry, we will use a staggered sublattice potential and introduce a weak heterostrain \cite{KTLAW_TBG_BCD,PacoTBGBCD,Arpitstraintbg}. This strained TBG/hBN hamiltonian can be described as 
\begin{equation}
    \hat H_{\xi}^{\rm TBG}(\mathbf{k})=
    \begin{pmatrix}
        \hat H_{b,\xi}(\mathbf{p}_{b,\xi}[\mathbf{k}]) & \hat U_\xi\\
        \hat U_\xi^\dagger & \hat H_{t,\xi}(\mathbf{p}_{t,\xi}[\mathbf{k}])
    \end{pmatrix},
\end{equation}
where $t,b$ are top and bottom layers at twist angle $\theta$ \cite{BMmodel,PhysRevX.8.031087}. Here $\xi =\pm1$ label the valleys and $U_\xi$ are the interlayer tunneling terms. In our calculations, we will sum both valley and spin degeneracy. Here $\mathbf{p}_{b,\xi}$ is a dimensionless momentum, see below. 

The tunneling matrices are $\hat U_\xi=U_1+U_2 e^{i\xi\mathbf G_{2}\cdot\mathbf r}+U_3 e^{i\xi\mathbf G_{1}\cdot\mathbf r}$ where the moir\'e reciprocal lattice vectors are $\mathbf G_1=\mathbf q_{tl}-\mathbf q_b$ and $\mathbf G_2=\mathbf q_{tr}-\mathbf q_b$. Here $\mathbf q_b =k_D \sin\frac\theta2 (0,-2),\mathbf q_{tr}=k_D\sin\frac\theta2  (\sqrt{3},1),\mathbf q_{tl}= k_D\sin\frac\theta2  (-\sqrt{3},1)$ \cite{KTLAW_TBG_BCD} are the generators of the Moir\'{e} lattice, $k_D = 4\pi/(3\sqrt{3} a_{\rm gr})$ and $U_1,U_2,U_3$ read as 
\begin{equation}
    U_1=\begin{pmatrix}
        u & u'\\
        u'&u
        \end{pmatrix},\qquad 
    U_2=\begin{pmatrix}
        u & u'e^{-i\xi2\pi/3}\\
        u'e^{i\xi2\pi/3} & u
        \end{pmatrix}, \qquad
    U_3 = \begin{pmatrix}
        u & u'e^{i\xi2\pi/3}\\
        u'e^{-i\xi2\pi/3} & u
        \end{pmatrix},
\end{equation}
$u=u'=110~{\rm meV}$ \cite{KTLAW_TBG_BCD}.

Each $t,b$ graphene layer is described by a Dirac Hamiltonian as 
\begin{equation}
    \hat H_{\ell,\xi}(\mathbf{p}_{\ell,\xi})=\alpha\left[p^x_{\ell,\xi}\,\xi\sigma_x+p^y_{\ell,\xi}\,\sigma_y\right]+\Delta_\ell^{\rm hBN}\sigma_z,\qquad \ell=b,t,
\end{equation}
where $\sigma_i$ are in the A/B sublattice space. Here $\Delta_\ell^{\rm hBN}$ describes the A/B sublattice breaking from stacking a hexagonal Boron Nitride onto the TBG. In our simulations we used $\alpha=2422.7~{\rm meV}$, $\Delta_b^{\rm hBN}=17~{\rm meV}$, and $\Delta_t^{\rm hBN}=0$ \cite{KTLAW_TBG_BCD}. 

\begin{figure*}
\centering
\includegraphics[width=0.99\textwidth]{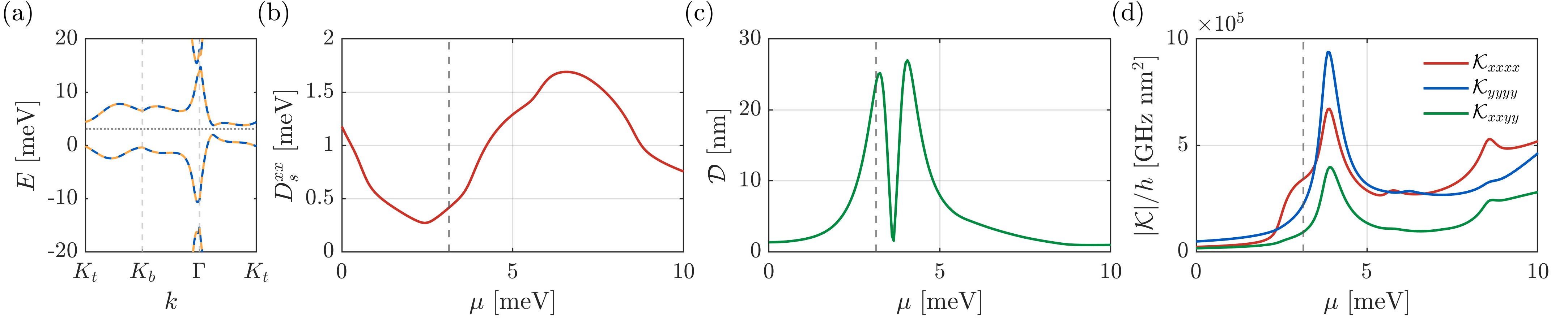}
\caption{
\textbf{Numerical estimate of superfluid weight, BCD, and Kerr effects in the superconducting proximitized twisted bilayer graphene (TBG) on hexagonal Boron Nitride platform.} Panel (a) shows the hBN-TBG band structure slice with moderate $\Delta_0=0.2$ meV used to generate all consequent plots of this figure. (b) is the superfluid weight as a function of chemical potential evaluated using Eq.~(\ref{generic_stiffness}). (c) the superconducting Berry curvature dipole evaluated using Eq.~(\ref{SBCDIBA}). (d) three intrinsic Kerr nonlinearities evaluated for the first band with $\epsilon_\mathbf{k}\geq0$ using Eq.~(\ref{Kfull}). Parameter values for the TBG/hBN calculation are listed in Section C of the Supplementary Materials.}\label{sBCDFIG2SI}
\end{figure*}

Heterostrain that breaks three-fold rotational symmetry is necessary for a finite Berry curvature dipole \cite{KTLAW_TBG_BCD,PacoTBGBCD}. In order to implement heterostrain, we shift the dimensionless momentas in opposite fashion on $t,b$ layers as \cite{KTLAW_TBG_BCD,PacoTBGBCD,Arpitstraintbg}
\begin{equation}
    \mathbf p_{b,\xi}[\mathbf{k}]=a_{\rm gr}~R(+\theta/2)\left[(1+\mathcal E_b^{\rm T})(\mathbf{k}-\mathbf K_{b,\xi})+\xi\mathbf A_b\right],\qquad
    \mathbf p_{t,\xi}[\mathbf{k}]=a_{\rm gr}~R(-\theta/2)\left[(1+\mathcal E_t^{\rm T})(\mathbf{k}-\mathbf K_{t,\xi})+\xi\mathbf A_t\right].
\end{equation}
In each layer, the Dirac points take on a form that accounts for both twist angle rotation as well as strain: for the bottom layer $\mathbf K_{b,\xi} =\xi R(-\theta/2)(1-\mathcal E_b^{\rm T})\mathbf K_0$ and for the top layer $\mathbf K_{t,\xi}=\xi R(+\theta/2)(1-\mathcal E_t^{\rm T})\mathbf K_0$~\cite{KTLAW_TBG_BCD,PacoTBGBCD,Arpitstraintbg} where $\mathbf K_0=-{(\mathbf b_1+\mathbf b_2)}/{3}$ are the usual bare Dirac points. Here $\mathbf b_1=2\pi\left(-1,-{1}/{\sqrt{3}}\right),\mathbf b_2=2\pi\left(-1,{1}/{\sqrt{3}}\right)$ are the bare graphene reciprocal lattice vectors. The uniaxial strain tensor is
\begin{equation}
    \mathcal E_\ell=\epsilon_\ell
    \begin{pmatrix}
    \cos^2\phi-\nu\sin^2\phi&(1+\nu)\cos\phi\sin\phi\\(1+\nu)\cos\phi\sin\phi&-\nu\cos^2\phi+\sin^2\phi
    \end{pmatrix},
\end{equation}
and the strain-induced gauge field is $\mathbf A=\sqrt{3}\,\beta_{\rm gr} (\mathcal E_{xx}-\mathcal E_{yy},-2\mathcal E_{xy})$. 
In our numerical simulations, we used a strained bottom layer and an unstrained top layer (realizing a heterostrain): $\epsilon_b=\epsilon$, $\epsilon_t=0$, $\mathbf A_b=\mathbf A$, and $\mathbf A_t=0$. Here we used numerical heterostrain parameters $\epsilon=0.001$, $\phi=0$, $\nu=0.165$, and $\beta_{\rm gr}=1.57$ \cite{KTLAW_TBG_BCD,PacoTBGBCD}. We also used twist angle $\theta=1.2^\circ$ and the graphene lattice constant $a_{\rm gr}=0.246~{\rm nm}$. In constructing the moiré supper lattice we truncated the plane-wave couplings at $N_{\rm moire}=5$, producing a large number of $4(2N_{\rm moire}+1)^2$ bands. We focused on the lowest narrow bands for the superfluid, BCD nonlinearity, and intrinsic Kerr nonlinearity calculations. We used a momentum grid of $500 \times 500$. 

TBG based superconductor simulations can be next performed by direct substitution of $\hat H_{b,\xi}(\mathbf{p}_{b,\xi}[\mathbf{k}])$ into the BdG Hamiltonian replacing $\hat H^{(0)}_{\mathbf{k}}$ in Eq.~(\ref{eq:BdGHSI}). Here we used the structureless order parameter $\hat \Delta = \mathbb{I} \cdot 0.2$ meV to simulate a proximity effect; we performed our calculation within the isolated band approximated (IBA) Hamiltonian as in Eq.~(\ref{reducedBdG}). Within IBA, we estimate the superfluid weight (see Eq.~(\ref{generic_stiffness}) and Fig.~\ref{sBCDFIG2SI}(b)), Berry curvature dipole (see Eq.~(\ref{SBCDIBA}) and Fig.~\ref{sBCDFIG2SI}(c)) and Kerr nonlinearities (see Eq.~(\ref{Kfull})) using the band dispersion and the wave function of the band that hosts the Fermi surface. 

We find that in proximitized TBG/hBN, BCD can be of order 20-30 nm and Kerr nonlinearites of order $\mathcal{K}_{xxxx}/h\sim 3\times 10^5$ [GHz nm$^{2}$] and $\mathcal{K}_{xxyy}/h\sim 10^5$ [GHz nm$^{2}$]. These values were used to illustrate the BCD quantum nonlinear circuit in the main text.
\end{document}